\documentclass[accept,a4paper]{mdpi} 

\usepackage{graphicx}
\usepackage{amssymb}
\usepackage{amsmath}
\usepackage{amscd}
\usepackage{eucal}
\usepackage{color}
\usepackage{bm}

\renewcommand{\today}{17 May 2016}

\firstpage{1} 
\articlenumber{208}
\doinum{10.3390/e18060208}
\pubvolume{18}
\pubyear{2016}
\copyrightyear{2016}
\externaleditor{Academic Editor: Ronnie Kosloff \  ({\it Entropy} special issue: {\it Quantum Thermodynamics})}
\history{Received: 30 March 2016; Revised: 17 May 2016; Published: 27 May 2016}

 \theoremstyle{mdpi}
 \newcounter{thm}
 \newcounter{ex}
 \newcounter{re}
 \theoremstyle{mdpidefinition}

\newcommand{\e}{{\rm e}}
\newcommand{\rmd}{{\rm d}}

\newcommand{\half}{{\textstyle{\frac{1}{2}}}}

\newcommand{\eps}{\epsilon}

\newcommand{\eminus}{ e^{\operatorname{-}} }
\newcommand{\kB}{k_{\rm B} }

\newcommand{\ignore}[1]{\relax}

\definecolor{DarkGreen}{rgb}{0,0.8,0}

\Title{QUANTUM COHERENT THREE-TERMINAL THERMOELECTRICS:
MAXIMUM EFFICIENCY AT GIVEN POWER OUTPUT}

\Author{Robert S.~Whitney}
\address{
Laboratoire de Physique et Mod\'elisation des Milieux Condens\'es (UMR 5493), 
Universit\'e de Grenoble  and CNRS, Maison des Magist\`eres, BP 166, 38042 Grenoble, France;
robert.whitney@grenoble.cnrs.fr}

\abstract{We consider the nonlinear scattering theory for three-terminal thermoelectric devices, 
used for power generation or refrigeration.
Such systems are quantum phase-coherent versions of a thermocouple, and the theory applies to
systems in which interactions can be treated at a mean-field level.
We consider an arbitrary three-terminal system in any external magnetic field, 
including systems with broken time-reversal symmetry, such as chiral 
thermoelectrics, as well as systems in which the magnetic field plays no role.  
We show that the upper bound on efficiency at given power output is of quantum origin and is stricter than Carnot's bound.
The bound is exactly the same as previously found for two-terminal devices, and can be achieved by three-terminal systems with or without broken time-reversal symmetry, i.e.~chiral and non-chiral thermoelectrics.}


\begin{document}

\section{Introduction}

Thermodynamics was the great product of nineteenth century physics;
it is epitomised by the concept that there is an upper bound on the efficiency of any thermodynamic machine, 
known as the Carnot limit.     This concept survived the quantum revolution with little more than a scratch; 
at present few physicists believe that a quantum machine can produce a significant amount of work at an efficiency 
exceeding the Carnot limit. Of course,
both statistical mechanics and quantum mechanics exhibit fluctuations, 
and these fluctuations may violate Carnot's limit on short timescales.  However, these fluctuations average out on longer timescales, so it is believed that any quantum machine left running long enough to produce a 
non-microscopic amount of work will not exceed the Carnot limit.
In this limit it is generally believed that Carnot's limit is only achievable for vanishing power output.
It was recently observed for  two-terminal thermoelectric machines that quantum mechanics imposes a {\it stricter} upper bound on the efficiency
at finite power output\cite{whitney-prl2014,whitney2015}.  This upper bound coincides with that
of Carnot at vanishing power output, but decays monotonically as one increases the desired power output.

\begin{figure}[t]
\centerline{\includegraphics[width=0.55\textwidth]{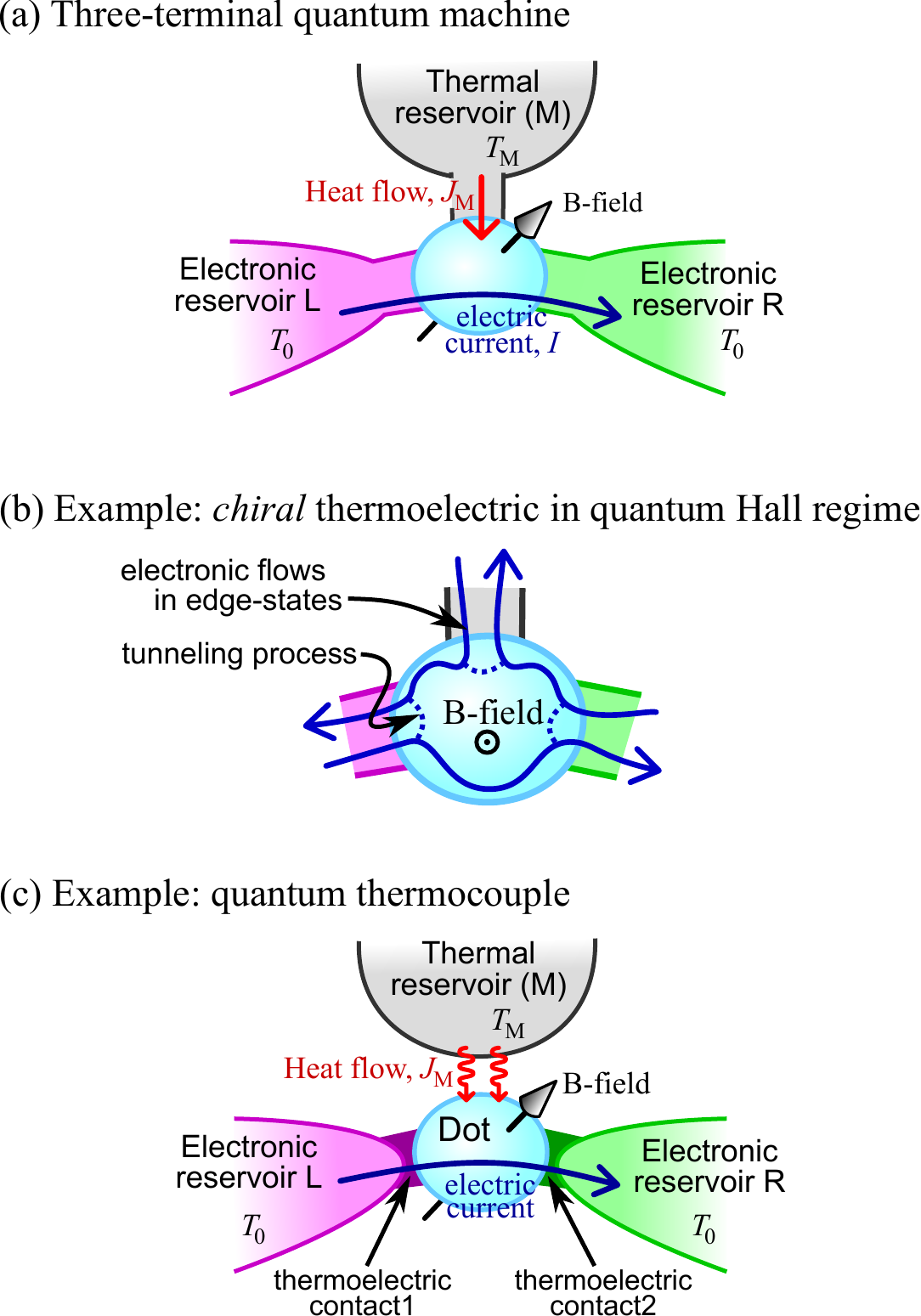}}
\caption{\label{Fig:three-term}
(a) The three-terminal machine (heat-engine or refrigerator) that we consider,  
the exchange of electrons with reservoir M carries a heat current, $J_{\rm M}$, but not an electrical current, $I_{\rm M}=0$.  
(b) A chiral thermoelectric device reproduced from Ref.~\cite{Sanchez2015a-qu-hall}.
(c) A system in which photons deliver the heat, this can be phenomenologically modelled by (a), see Section~\ref{Sect:voltage-probe}. 
}
\end{figure}

In recent years, there has been a lot of theoretical
\cite{Entin2010,Sanchez2011,Sothmann2012a,Entin2012,Jiang2012,Horvat2012,Brandner2013,Balachandran2013,
Jiang2013,Entin-Wohlman2013,Sanchez2013,Mazza2014,Jiang2014,Mazza2015,Sothmann2015,Sanchez2015a-qu-hall,Sanchez2015b-qu-hall,Jiang2015} and experimental \cite{Roche15,Hartmann15,Thierschmann15}
interest in three-terminal thermoelectrics, see Fig.~\ref{Fig:three-term}.
In particular, it is suggested that chiral three-terminal thermoelectrics \cite{Sanchez2015a-qu-hall,Sanchez2015b-qu-hall,Jiang2015} could have properties of great interest for efficient power generation.
Most of these three-terminal systems are quantum versions of traditional thermocouples 
\cite{books,DiSalvo-review,Shakouri-reviews},
since they have one terminal in contact with a thermal reservoir and two terminals in contact with electronic reservoirs. see Fig.~\ref{Fig:three-term}.  They turn heat flow from the thermal reservoir into electrical power in the electronic reservoirs, or vice versa.
We refer to such three-terminal systems as {\it quantum thermocouples}, since they are 
too small to be treated with the usual Boltzmann transport theory.  
There are two quantum lengthscales which enter into consideration; 
the electron's wavelength and its decoherence length.    In this work we will be interested in devices in which 
the whole thermocouple is much smaller than the decoherence length \cite{Brandner2013,Balachandran2013,Mazza2014,Sanchez2015a-qu-hall,Sothmann2015,Mazza2015,Sanchez2015b-qu-hall}.  Such thermocouples would typically be larger than the electron wavelength, although they need not be. The crucial point is that electrons flow elastically (without changing energy or thermalizing) through the central region in  Fig.~\ref{Fig:three-term}a.  
This can also be a simple phenomenological model of the system in Fig.~\ref{Fig:three-term}c, 
see Section~\ref{Sect:voltage-probe}.
In these systems, quantum interference effects can have a crucial effect on the physics.  
Such phase-coherent transport effects are not captured by the usual Boltzmann transport theory, 
but  they
can be modelled using Christen and B\"uttiker's nonlinear scattering theory \cite{Christen-Buttiker1996a}, in the cases where it is acceptable to treat electron-electron interactions 
at the mean-field level.
Such three-terminal systems are about the simplest self-contained quantum machines.

Reservoir M is taken to supply heat to the system but not electrical current. So 
the heat current into the system from reservoir M ($J_{\rm M}$) is finite,
while the electrical current into the system from reservoir M obeys 
\begin{eqnarray}
I_{\rm M}=0\, ,
\label{Eq:IM=0}
\end{eqnarray}
see Fig.~\ref{Fig:phenomenological}.
If reservoir L and R are at the same temperature $T_0$, and reservoir M is hotter at $T_{\rm M} > T_0$, 
we can use the heat flow $J_{\rm M}$ to drive an electrical current 
from L to R. If this electrical current flows against a potential difference, then the system turns heat into 
electrical power, and so is acting as a thermodynamic {\it heat-engine}.
Alternatively, we can make the system act as a {\it refrigerator},
by applying a bias which drives a current from L to R, and  ``sucks'' heat 
out of a reservoir M (Peltier cooling) taking it to a lower temperature than reservoirs L and R, $T_{\rm M} < T_0$.   

In this work, we consider arbitrary phase-coherent three-terminal quantum systems
that fall in to the category described by Christen and Buttiker's nonlinear scattering theory \cite{Christen-Buttiker1996a}.
We find upper bounds on such a system's efficiency as a heat-engine or a  refrigerator at finite power output.
We will show that these bounds coincide with those of two-terminal quantum systems considered in 
Ref.~\cite{whitney-prl2014,whitney2015}, irrespective of whether the three-terminal system's time-reversal symmetry is broken (by an external magnetic field) or not.  Thus our bound applies equally to normal and 
{\it chiral} thermoelectrics \cite{Sanchez2015a-qu-hall,Sanchez2015b-qu-hall,Jiang2015}.

\subsection{The Carnot bound}

When the system acts as a heat-engine (or 
energy-harvester \cite{Jordan-Sothmann-Sanchez-Buttiker2013,Sothmann-Sanchez-Jordan-Buttiker2013}),
the input is the heat current coming from the thermal reservoir (reservoir M), $J_{\rm M}$, 
and the output is the electrical power generated by the system,
$P_{\rm gen}$.  This power flows into a load attached between reservoirs L and R;
this load could be a motor turning electrical work into mechanical work, or some sort of work storage device.
The heat-engine (eng) efficiency is defined as
\begin{eqnarray}
\eta_{\rm eng} = P_{\rm gen}\big/ J_{\rm M}.
\label{Eq:eff-eng}
\end{eqnarray}  
This never exceeds Carnot's limit,
\begin{eqnarray}
\eta_{\rm eng}^{\rm Carnot} &=& 1-T_0/ T_{\rm M},
\label{Eq:Carnot-eng}
\end{eqnarray}
where we recall that $T_{\rm M} > T_0$.
For the refrigerator the situation is reversed, the load is replaced by a power supply, and the system absorbs power, $P_{\rm abs}$,  from that supply. The cooling power output is the heat current that is ``sucked''  out of the colder reservoir (reservoir M), $J_{\rm M}$.  
Thus the refrigerator (fri) efficiency or {\it coefficient of performance} (COP) is, 
\begin{eqnarray}
\eta_{\rm fri} = J_{\rm M} \big/ P_{\rm abs}.
\label{Eq:eff-fri}
\end{eqnarray}
This never exceeds Carnot's limit,
\begin{eqnarray}
\eta_{\rm fri}^{\rm Carnot} &=& (T_0/T_{\rm M} -1)^{-1}, \ \  
\label{Eq:Carnot-fri}
\end{eqnarray}
where we have $T_{\rm M} < T_0$ (which is the opposite of heat-engine). 

These Carnot limits are the upper bound on efficiency of heat-engines and refrigerators.
It has often been argued that Carnot efficiency is only achievable at zero cooling power, 
but no general proof of this claim exists, see Section~\ref{Sect:literature}.

\subsection{Stricter upper bound for two-terminal systems}

Bekenstein \cite{Bekenstein} and Pendry  \cite{Pendry1983} independently noted that there is an upper bound on the heat that can flow through a single transverse mode.  As a result, the heat that any wave (electron, photon, etc) can carry away from reservoir $i$ at temperature $T_i$ 
through a cross-section carrying $N$  transverse modes is
\begin{eqnarray}
J^{\rm qb}_{\rm i} = {\pi^2 \over 6 h}\  N \ \kB ^2 T_i^2
\end{eqnarray}
where the number of transverse modes is of order the cross-section in units of the wavelength of the particles carrying the heat.
  This Bekenstein-Pendry bound was observed experimentally in point-contacts \cite{Molenkamp-Peltier-thermalcond}, and recently verified to high accuracy in quantum Hall edge-states \cite{Jezouin2013}.  

Refs.~\cite{whitney-prl2014,whitney2015} pointed out that this upper bound on heat flow, must place a similar upper bound on the power generated by a heat-engine (since the efficiency is always finite). 
Those works used the nonlinear version of Landauer scattering theory \cite{Christen-Buttiker1996a} to find this upper bound on the power generated, which they called the quantum bound (qb), since its originates from the wavelike nature of electrons in quantum mechanics.  It takes the form
\begin{eqnarray}
P_{\rm gen}^{\rm qb} \,\equiv\, 
 A_0\, {\pi^2 \over h} \ N\  \kB^2 \big(T_L-T_R\big)^2, \quad \quad
\end{eqnarray}
where $A_0 \simeq 0.0321$. 
Refs.~\cite{whitney-prl2014,whitney2015} then calculated the upper bound on an heat engines efficiency for 
given power generation $P_{\rm gen}$ and showed that it is a monotonically decaying function of
$P_{\rm gen}\big/ P_{\rm gen}^{\rm qb}$.
There is no closed form algebraic expression for this upper bound at arbitrary $P_{\rm gen}\big/ P_{\rm gen}^{\rm qb}$, it is given by the solution of a transcendental equation. However, for $P_{\rm gen}\big/ P_{\rm gen}^{\rm qb}\ll 1$, 
the maximum efficiency at power $P_{\rm gen}$ is 
\begin{eqnarray}
\eta_{\rm eng} \big(P_{\rm gen}\big) =  \eta_{\rm eng}^{\rm Carnot} 
\left(1- 0.478
\sqrt{  {T_R \over T_L} \ {P_{\rm gen} \over P_{\rm gen}^{\rm qb}}} \,+ {\cal O} \left[P_{\rm gen} \big/ P_{\rm gen}^{\rm qb} \right]
\right)\! . \quad
\label{Eq:eta-eng-small-Pgen}
\end{eqnarray}
Thus one can only achieve Carnot efficiency at vanishing power generation, $P_{\rm gen}\to 0$,
although one comes close to Carnot efficiency for $P_{\rm gen}\ll P_{\rm gen}^{\rm qb}$.

In the limit of maximum power generation,  $P_{\rm gen}=P_{\rm gen}^{\rm qb}$, 
the upper bound on efficiency is 
\begin{eqnarray}
\eta_{\rm eng} (P_{\rm gen}^{\rm qb}) 
&=& {\eta_{\rm eng}^{\rm Carnot}\over 1+0.936 (1+T_R/T_L) }.
\label{Eq:Eff-at-Pqb}
\end{eqnarray}
Refs.~\cite{whitney-prl2014,whitney2015} calculated similar expressions for the upper bound on refrigerator efficiency as a function of cooling power.  In this case, the upper bound is found to be half the Bekenstein-Pendry bound on heat-flow.  Again, the maximum efficiency equals that of Carnot for cooling powers much less than the 
Bekenstein-Pendry bound, and decays monotonically as one increases the desired cooling power towards its upper limit.

In the naive classical limit of vanishing wavelength compared to system size, one has $N \to \infty$ and so the quantum bound 
$P_{\rm gen}^{\rm qb}$ and $J^{\rm qb}_i$ become irrelevant (they go to infinity).  So in this limit, it appears that one can achieve Carnot efficiency for any power output.  However, quantum mechanics says that this is not the case,
that for any power output that is a significant fraction of $P_{\rm gen}^{\rm qb}$ or $J^{\rm qb}_i$, the upper bound on efficiency is lower than Carnot efficiency.
This efficiency bound was derived for two-terminal quantum systems,
here we will show that exactly the same bounds apply to  three-terminal  quantum systems.

\begin{figure}
\centerline{\includegraphics[width=0.35\textwidth]{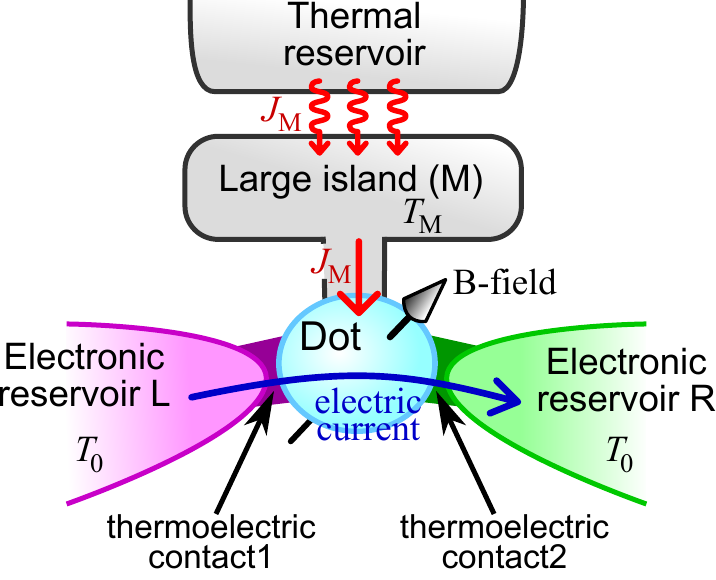}}
\caption{\label{Fig:phenomenological}
A sketch of a system for which the voltage-probe model discussed in Section~\ref{Sect:voltage-probe} is correct. 
The role of reservoir M is played by the island which is large enough that any electron entering it thermalizes at temperature $T_{\rm M}$ before escaping back into the dot.  The electro-neutrality of the island ensures that $I_{\rm M}=0$ in the steady-state.  However, the fact the island exchanges heat (in the form of photons or phonons)
with a thermal reservoir, means that it can still deliver heat to the three-terminal system.
The island is in a steady-state at temperature $T_{\rm M}$, for which the heat flow out of the island due to 
electrons, $J_{\rm M}$, equals the heat flow into the island due to photons (or phonons).
}
\end{figure}

\subsection{Universality of this bound? --- a brief literature review}
\label{Sect:literature}

The upper bound on efficiency at given power has been of some interest recently.
Various results have been derived in various regimes, the complexity of these calculations means that
there is not yet a consensus about how to compare these results. Here we attempt such a comparison, 
taking the risk that we may have misunderstood some of these complexities.

Many textbooks on thermodynamics give some sort of handwaving argument saying that
a heat-engine exhibiting Carnot efficiency has a vanishing power output, but this is by no means proven.  
In the specific context of the Carnot cycle, a step in this direction was 
made in the the pedagogical work of Curzon and Ahlborn \cite{Curzon-Ahlborn1975}
(although their result was found earlier \cite{Yvon1956,Chambadal57,Novikov57}),
which gave a curve for the efficiency as a function of the power of the machine,
and discussed in detail the efficiency at that machine's maximum power, $P_{\rm m}$.
In the linear response regime, one can use Onsager's non-equilibrium thermodynamics to show that this curve is particularly simple, it takes the form \cite{Casati-review}
 $\eta(P_{\rm gen}) = \half \eta_{\rm eng}^{\rm Carnot} \big(1+\sqrt{P_{\rm gen}/P_{\rm m}}\big)$, 
 see also Refs.~\cite{Horvat2012,Brandner2013,Balachandran2013,Benenti-et-al2011}.
 This goes linearly with $P_{\rm gen}$ when $P_{\rm gen}$ is small, rather than like a squareroot as in Eq.~(\ref{Eq:eta-eng-small-Pgen}).  However, a much bigger difference is that the theory does not give a value for $P_{\rm m}$, nor does it give an upper bound on $P_{\rm m}$.
As a result, such relations imply that  one could get arbitrarily close to Carnot efficiency at any finite power, $P_{\rm gen}$, by building a machine with $P_{\rm m}\to \infty$.
 For refrigerators, Ref.~\cite{Entin-Jiang-Imry2014}
showed that the entropy production rate goes like the power squared
with a prefactor that goes like $L_{\rm qq}\big/L_{\rho {\rm q}}^2$, 
where $L_{\mu\nu}$ is an Onsager coefficient with $\mu,\nu \in \rho$(charge)$,q$(heat). 
However, without a lower bound on $L_{\rm qq}\big/L_{\rho {\rm q}}^2$ (which may be power dependent),
this does not give us a lower bound on the entropy production rate at given refrigerator power
(such a lower bound would correspond to an upper bound on efficiency via Eq.~(\ref{Eq:dotS-fri})).  
We believe that it is quantum mechanics that gives the upper bounds on $P_{\rm m}$ (and on lower bound on $L_{\rm qq}\big/L_{\rho {\rm q}}^2$),
and so it is absent from these classical theories.

The first results which indicated the importance of quantum mechanics, were those that used scattering theory to  
show that Carnot efficiency required vanishingly narrow transmission functions
in both the linear \cite{Mahan-Sofo1996} and nonlinear regimes \cite{Humphry-Newbury-Taylor-Linke2002,Humphrey-Linke2005}
(Ref.~\cite{Mahan-Sofo1996} actually used Boltzmann transport theory, but every step of their calculation can be recast 
in terms of scattering theory if desired).  
A natural consequence of a vanishingly narrow transmission function is that the proportion of electrons that transmit through the thermoelectric structure is vanishing small.   This implies that the power output of such a system is vanishingly small for such a system\footnote{More strictly, this power output vanishes for any {\it finite sized} system (with a finite number of transverse modes). Formally, one can get a finite power in the limit, if one allows the machine's cross-section to diverge as one takes the transmission function's width to zero,  but this seems an unphysical way of taking the limit.}, irrespective of the bias one chooses. 
In the linear-response language these works tell us that the system whose figure of merit $ZT\equiv GS^2T\big/K\to \infty$ (Carnot efficiency requires $ZT\to \infty$),
has Onsager coefficients $L_{\mu\nu}$ whose magnitude's vanishes for all $\mu,\nu$, 
while the Seebeck coefficient $S \propto L_{\rho {\rm q}}\big/L_{\rho\rho}$ remain finite, and the Weidemann-Franz ratio
$K/(GT)\propto \big(L_{\rho \rho} L_{\rm qq} - L_{\rho {\rm q}}L_{{\rm q}\rho}\big)\big/ L_{\rho\rho}^2$ vanishes.  
Even if one chooses the load to maximize the power output, giving a power $P_{\rm m}$, the fact the transmission function is vanishingly narrow means that $P_{\rm m}\to0$.
This was the first indication that one could not take a machine's $P_{\rm m}$ to be independent of its efficiency.

This brings us to the scattering theory calculation in Refs.~\cite{whitney-prl2014,whitney2015}, outlined in the previous section.
There Eq.~(\ref{Eq:eta-eng-small-Pgen}) 
shows that Carnot efficiency is not achievable unless the power output is vanishing,
and that the deviation from Carnot efficiency goes like the squareroot of power.
This result is primarily for fully coherent transport, but 
Ref.~\cite{whitney2015} also considered a relaxation process within the scatterer as modelled by a fictitious reservoir
(in the style of a voltage probe \cite{Enquist-Anderson1981,Buttiker1986-4probe,voltage-probe,Imry-book}) in the absence of an external magnetic field, and recovered Eq.~(\ref{Eq:eta-eng-small-Pgen}) for small $P_{\rm gen}$.    Thus a natural question is how universal these bounds are.
The objective of this work is to show that the bounds in Refs.~\cite{whitney-prl2014,whitney2015} also applies to those relaxation-free three-terminal systems
which can be modelled by scattering theory.

However, returning to the effect of relaxation in two-terminal systems,
 Ref.~\cite{Brandner2015} considered the linear-response limit of scattering theory
with an arbitrary number of fictitious reservoirs to model more complicated relaxation processes (and with an arbitrary external magnetic field).  
They considered maximizing the power for given efficiency,
and found a bound that was weakest when the number of fictitious reservoirs
goes to infinity.  In the limit of small power, their result gives the maximum efficiency for given power as
\begin{eqnarray}
\eta_{\rm eng}(P) = \eta_{\rm eng}^{\rm Carnot} \times \big(1- P_{\rm gen}/(4P_0) + \cdots \big)
\end{eqnarray}
where $P_0$ is the same as $P_{\rm gen}^{\rm max}$, except for a difference in the numerical prefactor.
The absence of the square-root makes this bound is much less strict at small $P_{\rm gen}$ 
than Eq.~(\ref{Eq:eta-eng-small-Pgen}). This hints that it might be possible to exceed 
Eq.~(\ref{Eq:eta-eng-small-Pgen}) 
by adding a large amount of relaxation within the scatterer (as modelled by an infinite number of fictitious reservoirs).  However, Ref.~\cite{Brandner2015} say that their upper bound may be an over-estimate; they do not prove it is a tight bound by giving an example of a system that achieves their upper bound.  Thus, we  cannot yet say 
for certain whether a system of the type that they propose can violate the bound in Eq.~(\ref{Eq:eta-eng-small-Pgen}) or not.
Similarly, nothing is know about the bound for systems which are not modelled by scattering theory,
such as systems exhibiting strong interaction effects (Coulomb-blockade, Kondo effect, etc).
So it remains to be seen how universal this bound is, even if Eq.~(\ref{Eq:eta-eng-small-Pgen}) is obeyed by 
all the systems for which a tight bound has been derived to date.

\subsection{Examples of three terminal systems: chiral thermoelectrics and quantum thermocouples}
\label{Sect:voltage-probe}

Here we discuss two examples of systems for which the bounds we derive here apply.
The first example is the {\it chiral thermoelectric} sketched in Fig.~\ref{Fig:three-term}b, 
as discussed in Refs.~\cite{Sanchez2015a-qu-hall,Sanchez2015b-qu-hall,Jiang2015}.  This is a three-teminal system exposed to such a strong external magnetic field that the electron flow only occurs via edge-states (all bulk states are localized by the magnetic field). These edge-state are chiral, which means they circulate in a preferred direction in the scattering region (anticlockwise in Fig.~\ref{Fig:three-term}b), 
This is an intriguing situation for a heat-engine in which one wants to generate electrical power by driving a flow of electrons from
reservoir L (at lower chemical potential) to reservoir R (at higher chemical potential). The B-field alone
generates an electron flow directly  from L to R {\it without} a corresponding direct electron flow from R to L.  
Thus it would seem plausible that one could take advantage of this, with a suitable choice of Reservoir M
and of the central scattering region to achieve higher efficiencies than in a two-terminal device (where every flow from
L to R  has a corresponding flow from R to L).
Unfortunately, our general solution for a three-terminal system will show that
the upper bound on efficiency at given power output is independent of the external magnetic field, so it is the same for chiral or non-chiral systems.  

The second example is the quantum thermocouple sketched in Fig.~\ref{Fig:three-term}c.
Here, the third terminal (reservoir M) supplies heat in the form of photons (or phonons).
Such systems have been considered using microscopic models of the photon flow 
\cite{Entin2010,Sanchez2011,Sothmann2012a,Entin2012,Jiang2012,
Jiang2013,Entin-Wohlman2013,Jiang2014,Jiang2015},
however here we instead use a phenomenological argument to replace the reservoir of photons sketched in
Fig.~\ref{Fig:three-term}c by the reservoir of electrons sketched in Fig.~\ref{Fig:three-term}a.
This is the ``voltage probe'' model \cite{Enquist-Anderson1981,Buttiker1986-4probe,voltage-probe,Imry-book}, in which inelastic scattering (such as electrons scattering from photons)
is modelled by a reservoir of electrons whose chemical potential is chosen such that 
that on average every electron that escapes the system into that reservoir is replaced by one coming into the system from that reservoir, so $I_{\rm M}=0$.
Fig.~\ref{Fig:phenomenological} shows a system for which this voltage probe model is correct.
The island is large enough that any electron entering it thermalizes at temperature $T_{\rm M}$ before escaping back into the dot.  Since the island is in a steady-state at temperature $T_{\rm M}$, the heat flow out of the island due to 
electrons must equal the heat flow into the island due to photons.
However, one can also argue phenomenologically that the same model is a simplified description of the system sketched in Fig.~\ref{Fig:three-term}c.
This phenomenological model treats the exchange of a photon between the dot and reservoir M, as the replacement of an electron in the dot which has the dot's energy distribution, with an electron which has reservoir M's energy distribution.
Of course, this is not the most realistic model of electron-photon interactions. In particular, it assumes that each electron entering from reservoir L or R either escapes into one of those two reservoirs without any inelastic scattering from the photon-field, or it escapes after undergoing so many scatterings from the photon-field that it has {\it completely} thermalized with the photon-field. As such, this model does not capture the physics of electrons that undergo one or two inelastic scatterings from the photon-field before escaping into reservoir L or R.  At this simplistic level of modelling, nothing would change if it were phonons rather than photons
coming from reservoir M.
While this voltage probe model has been successfully used to understand the basics of many 
inelastic effects in nanostructures, it should not be considered a replacement for a proper microscopic theory 
(see e.g.\ Refs.~\cite{pjw2007,wjp2008} for a discussion of how the voltage probe model fails to capture
aspects of inelastic scattering in ultra-clean nanostructures).
One should be cautious about applying results for a system of the type in Fig.~\ref{Fig:three-term}a
to a system of the type in Fig.~\ref{Fig:three-term}c, but it is none the less a reasonable first step to understanding
its physics.

\section{Electrical and heat currents}

Consider a system with a scattering matrix,  ${\cal S}(\eps)$,
then the transmission matrix for electrons at energy $\eps$ made of elements 
\begin{eqnarray}
{\cal T}_{ij}(\eps) ={\rm tr} \left[ {\cal S}_{ij}^\dagger (\eps) {\cal S}_{ij}(\eps) \right],
\label{Eq:def-T}
\end{eqnarray} 
where the trace is over all transverse modes of leads $i$ and $j$.
The  electrical current out of reservoir $i$ is then 
\begin{eqnarray}
I_i = \eminus \int_{-\infty}^\infty {\rmd \eps \over h}  \sum_j \Big( {\cal T}_{ij}(\eps)  -N_i(\eps) \delta_{ij}\Big) f_j(\eps),
\end{eqnarray}
where lead $i$ has $N_i (\eps)$ modes for particles at energy $\eps$,
and we define the Fermi function in reservoir $j$ as
\begin{eqnarray}
f_j(\eps) = \left( 1+\exp\big [(\eps-\eminus V_j)\big/(\kB T_j) \big] \right)^{-1}.
\label{Eq:Fermi}
\end{eqnarray}
The heat-current out of reservoir $i$ is
\begin{eqnarray}
J_i = \int_{-\infty}^\infty {\rmd \eps \over h} (\eps -\eminus V_i) \sum_j  \Big( {\cal T}_{ij}(\eps)  -N_i(\eps) \delta_{ij}\Big)  f_j(\eps).\ 
\end{eqnarray}
The unitarity of ${\cal S}$ places the following constraints on 
the transmission functions.
Firstly,
\begin{eqnarray}
N_i(\eps)  = \sum_j  {\cal T}_{ij}(\eps) = \sum_j {\cal T}_{ji}(\eps) \, .
\label{Eq:constraint1}
\end{eqnarray}
Secondly,
\begin{eqnarray}
0 \ \leq \ &{\cal T}_{ij}(\eps)& \ \leq \ N^{\rm min}_{ij}  \ .
\label{Eq:constraint2}
\end{eqnarray}
where for compactness in what follows we define
\begin{eqnarray}
N^{\rm min}_{ij} = {\rm min}[N_i,N_j]\ .
\label{Eq:Nmin}
\end{eqnarray}
It has been shown \cite{Nenciu2007,whitney2013} that any three-termnal system obeying the above theory automatically satisfies the laws of thermodynamics, if one takes  the Clausius definition of entropy for the reservoirs. This means that
the rate of entropy production is $-\sum_i J_i\big/T_i$, where the sum is over all reservoirs.

\subsection{Currents for three-terminal systems}

A system with three terminals has a three-by-three transmission matrix, meaning it has 
nine transmission functions. However, Eq.~(\ref{Eq:constraint1}) means that only 
four of them are {\it independent}.
There are many possible choices for these four, we choose
\begin{eqnarray}
 {\cal T}_{\rm LM}(\eps), \ {\cal T}_{\rm RM}(\eps), \ {\cal T}_{\rm LR}(\eps), \hbox{ and } {\cal T}_{\rm RL}(\eps).
\label{Eq:Tmatrix-4elements}
\end{eqnarray}
The remaining five transmission functions are written in terms of these functions;
\begin{subequations}
\label{Eq:5othertransmissions}
\begin{eqnarray}
{\cal T}_{\rm LL}(\eps) &=& N_{\rm L}(\eps)-{\cal T}_{\rm LM}(\eps) -{\cal T}_{\rm LR}(\eps) , \\
{\cal T}_{\rm RR}(\eps) &=& N_{\rm R}(\eps) -{\cal T}_{\rm RL}(\eps)   - {\cal T}_{\rm RM}(\eps)  , \\
{\cal T}_{\rm MM}(\eps) &=& N_{\rm M}(\eps)-{\cal T}_{\rm LM}(\eps) - {\cal T}_{\rm RM}(\eps) ,  \\
{\cal T}_{\rm ML}(\eps) &=&{\cal T}_{\rm LM}(\eps) +{\cal T}_{\rm LR}(\eps) -{\cal T}_{\rm RL}(\eps) ,
\label{Eq:T_ML}\\
{\cal T}_{\rm MR}(\eps) &=& {\cal T}_{\rm RM}(\eps)+{\cal T}_{\rm RL}(\eps)-{\cal T}_{\rm LR}(\eps) .
\label{Eq:T_MR}
\end{eqnarray}
\end{subequations}

\begin{figure}[t]
\centerline{\includegraphics[width=\textwidth]{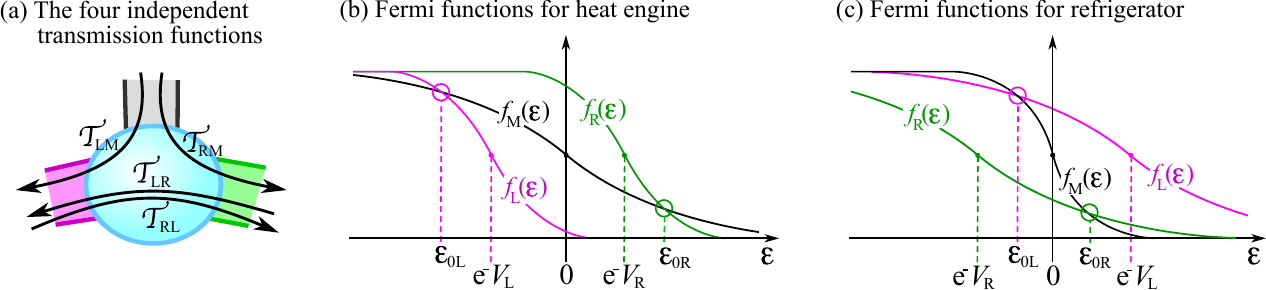}}
\caption{\label{Fig:fermi-functions}
In (a) we sketch the four transmission functions which completely determine the system's scattering properties,
it is particle conservation that enables us to completely determines the remaining five transmission and reflection processes from these four, see Eqs.~(\ref{Eq:5othertransmissions}). 
In (b) and (c) we sketch the Fermi-functions for each reservoir for the case of a heat-engine and refrigerator, respectively.
For the heat-engine we have $T_{\rm L}=T_{\rm R} < T_{\rm M}$ and $\eminus V_{\rm L} < 0 < \eminus V_{\rm R}$,
as discussed in Section~\ref{Sect:eng}.
For the refrigerator we have $T_{\rm L}=T_{\rm R} > T_{\rm M}$ and $\eminus V_{\rm L} > 0 > \eminus V_{\rm R}$,
as discussed in Section~\ref{Sect:fri}.
}
\end{figure}

Given these relations between transmission matrix elements, we can write currents into the quantum system from reservoirs ($L,R,M$) as
\begin{eqnarray}
I_{\rm L} &=& \eminus \int_{-\infty}^\infty {\rmd \eps \over h}\  \Big(
{\cal T}_{\rm LM}(\eps) \,\big[f_{\rm L}(\eps) -f_{\rm M}(\eps)\big] 
\ +\ {\cal T}_{\rm LR}(\eps) \,\big[f_{\rm L}(\eps) -f_{\rm R}(\eps)\big] \Big),
\label{Eq:I_L-integral}
\\
I_{\rm R} &=& \eminus \int_{-\infty}^\infty {\rmd \eps \over h}\  \Big(
{\cal T}_{\rm RM}(\eps) \,\big[f_{\rm R}(\eps) -f_{\rm M}(\eps)\big] 
\ +\ {\cal T}_{\rm RL}(\eps) \,\big[f_{\rm R}(\eps) -f_{\rm L}(\eps)\big] \Big),
\label{Eq:I_R-integral}
\\
I_{\rm M} &=& -I_{\rm L}-I_{\rm R} ,
\label{Eq:I_H-integral}
\end{eqnarray}
We chose to measure chemical potentials from that of reservoir M, so $V_{\rm M}=0$.
Then the heat current out of reservoir M is
\begin{eqnarray}
J_{\rm M} &=& 
\int_{-\infty}^\infty {\eps \,\rmd \eps \over h}\ \Big(
\big[{\cal T}_{\rm LR}(\eps) - {\cal T}_{\rm RL}(\eps)\big] \,\big[f_{\rm R}(\eps) -f_{\rm L}(\eps)\big]
+ \,{\cal T}_{\rm LM}(\eps) \,\big[f_{\rm M}(\eps) -f_{\rm L}(\eps)\big] 
\nonumber \\
& & \qquad \qquad \qquad
+ \, {\cal T}_{\rm RM}(\eps) \,\big[f_{\rm M}(\eps) -f_{\rm R}(\eps)\big] \Big). 
\label{Eq:J_M-integral}
\end{eqnarray}
The power generated is 
\begin{eqnarray}
P_{\rm gen} &=& -V_{\rm L}I_{\rm L} - V_{\rm R}I_{\rm R} \, .
\label{Eq:Pgen-integral}
\end{eqnarray}

\section{Transmission which maximizes heat engine efficiency for given power output}
\label{Sect:eng}

Our objective is to find the transmission functions,  ${\cal T}_{\rm LM}(\eps)$, ${\cal T}_{\rm RM}(\eps)$, 
${\cal T}_{\rm LR}(\eps)$, and ${\cal T}_{\rm RL}(\eps)$, that maximize the heat-engine efficiency for given power 
generation, $P_{\rm gen}$.  This is equivalent to finding the transmission functions that minimize heat flow out of reservoir M, $J_{\rm M}$, for given $P_{\rm gen}$.
To find these optimal transmission functions we must start with completely arbitrary $\eps$ dependences of the transmission functions.
As in Refs.~\cite{whitney-prl2014,whitney2015}, we do this by considering each transmission function as consisting of an infinite number of slices, each of vanishing width $\delta$.
We define  $\tau^{(\gamma)}_{ij}$ as the height of the $\gamma$th slice of ${\cal T}_{ij}(\eps)$, which is the slice with energy $\eps_\gamma$. 
We then want to optimize the biases of reservoirs L and R ($V_{\rm L}$ and $V_{\rm R}$)
and each $\tau^{(\gamma)}_{ij}$; this requires finding the value of each of this infinite number of parameters
that minimize $J_{\rm M}$ under the constraints that $I_{\rm M}=0$ and that $P_{\rm gen}$ is fixed at the value of interest.

The central ingredients in this optimization are the rate of change of 
$P_{\rm gen}$, $I_{\rm M}$ and $J_{\rm M}$ with $\tau^{(\gamma)}_{ij}$.  
Here, 
\begin{eqnarray}
{\rmd P_{\rm gen} \over  \rmd \tau^{(\gamma)}_{ij} }\Bigg|_{V,\tau}
&=& \eminus V_i  \, {\delta \over h}\,\big[f_j(\eps_\gamma) -f_i(\eps_\gamma)\big] ,
\label{Eq:dPgen/dtau}
\end{eqnarray}
where $\big|_{V,\tau}$ means the derivative is taken for fixed $V_{\rm L}$, $V_{\rm R}$ and fixed $\tau^{(\gamma')}_{ij}$
for all  $\gamma' \neq\gamma$.
Doing the same for $I_{\rm M}$ and $J_{\rm M}$, we get
for $ij \in \{ {\rm LM,RM,LR,RL} \}$, 
\begin{eqnarray}
{\rmd I_{\rm M} \over \rmd  \tau^{(\gamma)}_{ij} }\Bigg|_{V,\tau}
&=& {1 \over V_i} {\rmd P_{\rm gen} \over \rmd  \tau^{(\gamma)}_{ij} }\Bigg|_{V,\tau}\ ,
\label{Eq:IH-to-Pgen}
\\
{\rmd J_{\rm M} \over  \rmd \tau^{(\gamma)}_{ij} }\Bigg|_{V,\tau}
&=& {\eps_\gamma \over \eminus V_i} {\rmd P_{\rm gen} \over \rmd  \tau^{(\gamma)}_{ij} }\Bigg|_{V,\tau}\ .
\label{Eq:JH-to-Pgen}
\end{eqnarray}

For a heat-engine, we consider the case where
$T_{\rm L}=T_{\rm R}=T_0$ and $T_{\rm M} > T_0$, while  $\eminus V_{\rm L} \, < \, 0\, < \, \eminus V_{\rm R}$.
The Fermi functions in this case are sketched in Fig.~\ref{Fig:fermi-functions}a.
We observe that
\begin{eqnarray}
\big[f_{\rm R}(\eps)-f_{\rm L}(\eps)\big]& \hbox{ is} &  \hbox{positive for all } \eps \ ,
\label{Eq:sign-of-F-difference2}
\\
\big[f_{\rm M}(\eps)-f_i(\eps)\big] & \hbox{ is} &
\left\{ \begin{array}{l}
\hbox{ positive for $\eps >\eps_{0i}$} \,,\\
\hbox{ negative for $\eps <\eps_{0i}$} \,,
\end{array}\right.
\label{Eq:sign-of-F-difference1}
\end{eqnarray}
where we define 
\begin{eqnarray}
\eps_{0i} = {\eminus V_i \over 1-T_0/T_{\rm M}}\ .
\label{Eq:eps0}
\end{eqnarray}
We will take $V_{\rm L},V_{\rm R}$ such that $I_{\rm M}=0$, and 
\begin{eqnarray}
I_{\rm R}=-I_{\rm L} >0   \ .
\end{eqnarray}
To proceed with the derivation it is more convenient to 
assume we are interested in minimizing the heat-flow $J_{\rm M}$ for given 
$P_{\rm gen}$ and given $I_{\rm M}$.  Only at the end will we take
$I_{\rm M}=0$, to arrive at the situation of interest.

\subsection{Optimizing 
${\cal T}_{\rm RM}$,  ${\cal T}_{\rm LM}$,  ${\cal T}_{\rm RL}$ and ${\cal T}_{\rm LR}$ independently}
\label{Sect:independent}

We start with the assumption that the four transmission functions, 
${\cal T}_{\rm RM}$,  ${\cal T}_{\rm LM}$,  ${\cal T}_{\rm RL}$ and ${\cal T}_{\rm LR}$,  
each have a completely arbitrary energy dependence, and can be optimized independently.
Only in section~\ref{Sect:constraints} do we take into account the relations between these transmission functions
imposed by combining Eq.~(\ref{Eq:constraint2}) with Eq.~(\ref{Eq:5othertransmissions}).

To carry out the independent optimization of each of the four transmission functions, 
let us define 
\begin{eqnarray}
\partial_{\rm R} \cdots ={\rmd (\cdots) \over \rmd V_{\rm R}}\Bigg|_{V_{\rm L},{\cal T}} \ , 
 \hbox{ and } \ 
\partial_{\rm L} \cdots ={\rmd (\cdots) \over \rmd V_{\rm L}}\Bigg|_{V_{\rm R},{\cal T}}  \ ,
\label{Eq:define-dL-and-dR}
\end{eqnarray}
where $|_{V_i,{\cal T}}$ indicates that the derivative is for fixed $V_i$ and fixed transmission functions.
Then, for an inifinitesimal change of $\tau^{(\gamma)}_{ij}$, $V_{\rm L}$ and $V_{\rm R}$ we have 
\begin{eqnarray}
\delta J_{\rm M} \!\!  &=& 
{\rmd J_{\rm M} \over \rmd \tau^{(\gamma)}_{ij}} \Bigg|_{V,\tau} \! \delta \tau^{(\gamma)}_{ij} 
+  \partial_{\rm L} J_{\rm M}  \delta V_{\rm L} + \partial_{\rm R} J_{\rm M}  \delta V_{\rm R} \, ,\qquad 
\label{Eq:deltaJ_H}
\\
\delta I_{\rm M} \!\!  &=& 
{\rmd I_{\rm M} \over \rmd \tau^{(\gamma)}_{ij}} \Bigg|_{V,\tau} \! \delta \tau^{(\gamma)}_{ij} 
 +  \partial_{\rm L} I_{\rm M}  \delta V_{\rm L} + \partial_{\rm R} I_{\rm M}  \delta V_{\rm R} \, ,\qquad 
\label{Eq:deltaI_H}
\\
\delta P_{\rm gen} \!\!\!  &=& 
{\rmd P_{\rm gen}  \over \rmd \tau^{(\gamma)}_{ij}} \Bigg|_{V,\tau} \! \delta \tau^{(\gamma)}_{ij} 
+  \partial_{\rm L}  P_{\rm gen}   \delta V_{\rm L} + \partial_{\rm R}  P_{\rm gen}   \delta V_{\rm R} \, .\qquad 
\label{Eq:deltaPgen}
\end{eqnarray}
We are interested in fixed $P_{\rm gen}$ and $I_{\rm M}$, so we want
$\delta I_{\rm M} = \delta P_{\rm gen}=0$.  This means Eqs.~(\ref{Eq:deltaI_H},\ref{Eq:deltaPgen}) 
form a pair of simultaneous equations, which we solve to get
\begin{eqnarray}
 \delta V_{\rm L} \!\! &=& \!\!  
 \left[ {\partial_{\rm R} I_{\rm M} \over {\cal A}}  
{\rmd P_{\rm gen}  \over \rmd \tau^{(\gamma)}_{ij}} \Bigg|_{V,\tau} 
\!-  {\partial_{\rm R} P_{\rm gen}  \over {\cal A}} 
{\rmd I_{\rm M} \over \rmd \tau^{(\gamma)}_{ij}} \Bigg|_{V,\tau} \right]  \, \delta \tau^{(\gamma)}_{ij}  ,
\nonumber 
\\
 \delta V_{\rm R} \!\! &=& \!\! 
\left[-{\partial_{\rm L} I_{\rm M}  \over {\cal A}}
 {\rmd P_{\rm gen}  \over \rmd \tau^{(\gamma)}_{ij}} \Bigg|_{V,\tau} 
 \!\!+  {\partial_{\rm L} P_{\rm gen}  \over {\cal A}} 
 {\rmd I_{\rm M} \over \rmd \tau^{(\gamma)}_{ij}} \Bigg|_{V,\tau} \right]   \delta \tau^{(\gamma)}_{ij}  ,
\nonumber
\end{eqnarray}
where we define 
\begin{eqnarray}
 {\cal A} &=& 
 \partial_{\rm L} I_{\rm M}\ \partial_{\rm R} P_{\rm gen} \ -\   \partial_{\rm R} I_{\rm M}\ \partial_{\rm L} P_{\rm gen}\ .
\nonumber
\end{eqnarray}
We substitute these results for $\delta V_{\rm L}$ and $\delta V_{\rm R}$ into Eq.~(\ref{Eq:deltaJ_H})
and use Eqs.~(\ref{Eq:IH-to-Pgen},\ref{Eq:JH-to-Pgen}) to cast everything in terms of   
$\rmd P_{\rm gen}  \big/ \rmd \tau^{(\gamma)}_{ij}$. 
Then for $ij \in \{ {\rm LM,RM,LR,RL} \}$, 
\begin{eqnarray}
\delta J_{\rm M} 
&=& 
  \delta \tau^{(\gamma)}_{ij} \ \left[
{\eps_\gamma -\eps_{1i} \over \eminus V_i}
 \right] \ {\rmd P_{\rm gen}  \over \rmd \tau^{(\gamma)}_{ij}} \Bigg|_{V,\tau}
\ ,\qquad 
\label{Eq:dJ_H-result}
\end{eqnarray}
where we define $\eps_{1i}$, with $i\in L,R$, as
\begin{eqnarray}
\eps_{1i}
&=&
\eminus V_i \ {\partial_{\rm R} J_{\rm M}  \ \partial_{\rm L} I_{\rm M} -\partial_{\rm L} J_{\rm M} \  \partial_{\rm R} I_{\rm M}   \over {\cal A}} 
\ + \ \eminus \ {\partial_{\rm L} J_{\rm M}  \ \partial_{\rm R} P_{\rm gen} - \partial_{\rm R} J_{\rm M}  \ \partial_{\rm L} P_{\rm gen}  \over {\cal A}} \ .
\label{Eq:eps1}
\end{eqnarray}
Thus, using Eq.~(\ref{Eq:dPgen/dtau}), we conclude that $J_{\rm M}$ shrinks upon increasing $\tau^{(\gamma)}_{ij}$ 
(for fixed $P_{\rm gen}$ and fixed $I_{\rm M}$) if
\begin{eqnarray}
 \left[
\eps_\gamma  - \eps_{1i} \right] \, \big[f_j(\eps_\gamma) -f_i(\eps_\gamma)\big] &<& 0 \ ,
\label{Eq:nearly-at-boxcars}
\end{eqnarray}
and otherwise $J_{\rm M}$ grows upon increasing $\tau^{(\gamma)}_{ij}$.
The sign of the difference of Fermi functions is given by
Eqs.~(\ref{Eq:sign-of-F-difference2},\ref{Eq:sign-of-F-difference1}). Hence,
$J_{\rm M}$ is reduced for fixed $P_{\rm gen}$ and fixed $I_{\rm M}$
by 
\begin{itemize}
\item[(a)] increasing ${\cal T}_{\rm RM}(\eps)$ up to $N^{\rm min}_{\rm RM}$ for $\eps$ between $\eps_{\rm 0R}$ and $\eps_{\rm 1R}$, while reducing 
 ${\cal T}_{\rm RM}(\eps)$ to zero for all other $\eps$.

\item[(b)]  increasing ${\cal T}_{\rm LM}(\eps)$ up to $N^{\rm min}_{\rm LM}$ for $\eps$ between $\eps_{\rm 0L}$ and $\eps_{\rm 1L}$, while reducing 
 ${\cal T}_{\rm LM}(\eps)$ to zero for all other $\eps$.

\item[(c)] increasing ${\cal T}_{\rm RL}(\eps)$ up to $N^{\rm min}_{\rm RL}$  for $\eps>\eps_{\rm 1R}$, while reducing 
 ${\cal T}_{\rm LM}(\eps)$ to zero for $\eps<\eps_{\rm 1R}$.

\item[(d)] increasing ${\cal T}_{\rm LR}(\eps)$ up to $N^{\rm min}_{\rm RL}$ for $\eps<\eps_{\rm 1L}$, while reducing 
 ${\cal T}_{\rm LM}(\eps)$ to zero for $\eps>\eps_{\rm 1L}$.
\end{itemize}
Here, it is Eq.~(\ref{Eq:constraint2}) that stops us reducing these functions
below zero, or increasing  ${\cal T}_{ij}(\eps)$ beyond $N^{\rm min}_{ij}$.

While it is hard to guess the form of $\eps_{\rm 1L}$ and $\eps_{\rm 1R}$ 
from their definition in Eq.~(\ref{Eq:eps1}).  By inspecting Eqs.~(\ref{Eq:I_L-integral},\ref{Eq:I_R-integral}) one sees that a heat-engine should have $\eps_{\rm 1R} > \eps_{\rm 0R}$ and $\eps_{\rm 1L}< \eps_{\rm 0L}$ to ensure that both terms contributing to $P_{\rm gen}$ in Eq.~(\ref{Eq:Pgen-integral}) are positive.
While refrigerators are not discussed until Section~\ref{Sect:fri}, we will show there that their optimization leads to similar rules to (a-d) above.  However, refrigerators must absorb electrical power (negative $P_{\rm gen}$), so they will have $\eps_{\rm 1R} < \eps_{\rm 0R}$ and $\eps_{\rm 1L}> \eps_{\rm 0L}$, with section~\ref{Sect:fri} also showing that $\eps_{\rm 1R}>0$ and $\eps_{\rm 1L}<0$.
Thus, we will consider two situations,
\begin{subequations}
\begin{eqnarray}
\hbox{heat-engine: } & &  \eps_{\rm 1L} < \eps_{\rm 0L} < 0 < \eps_{\rm 0R} < \eps_{\rm 1R}, \qquad
\label{Eq:eng-eps-ordering}
\\
\hbox{refrigerator: } & &  \eps_{\rm 0L} < \eps_{\rm 1L} < 0 < \eps_{\rm 1R} < \eps_{\rm 0R}.
\label{Eq:fri-eps-ordering}
\end{eqnarray}
\end{subequations}
as sketched in Fig.~\ref{Fig:boxcars}a and \ref{Fig:boxcars}b, respectively.

\subsubsection{{\it Problem with the independent optimization}}

\begin{figure}[t]
\centerline{\includegraphics[width=\textwidth]{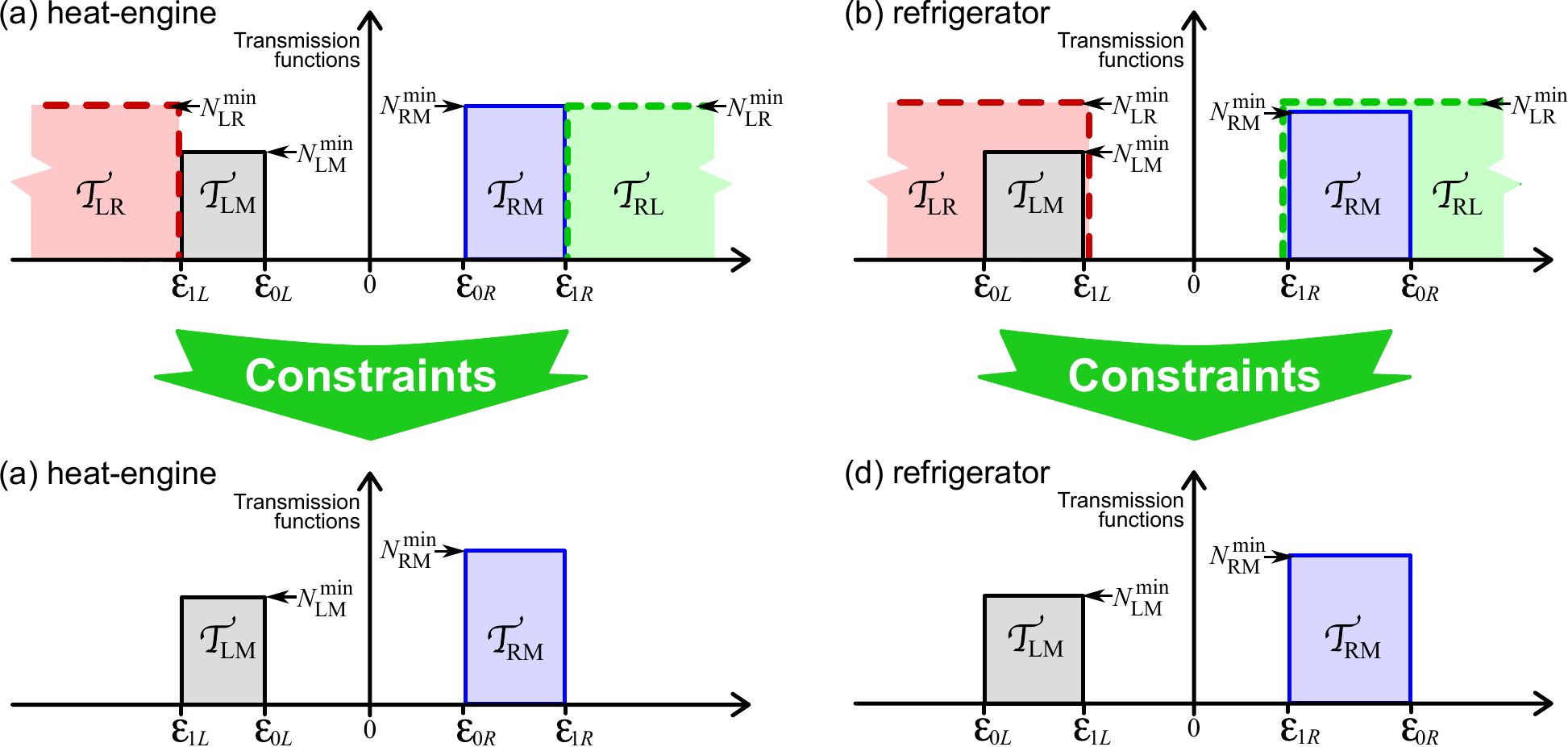}}
\caption{\label{Fig:boxcars}
If we could maximize
${\cal T}_{\rm RM}$,  ${\cal T}_{\rm LM}$,  ${\cal T}_{\rm RL}$ and ${\cal T}_{\rm LR}$ independently,
as discussed in section~\ref{Sect:independent}, we would get the optimal boxcar functions like those in (a) or (b). 
The height of the boxcar for ${\cal T}_{ij}$ is
$N^{\rm min}_{ij}$ defined in Eq.~(\ref{Eq:Nmin}), for concreteness in the sketch we take $N_{\rm R} < N_{\rm M} <N_{\rm L}$.
Once we introduce the constraints discussed section~\ref{Sect:constraints}, we get the boxcar functions in (c) or (d), given by Eqs.~(\ref{Eq:final-Ts}) and Eqs.~(\ref{Eq:fri-final-Ts}), respectively. 
}
\end{figure}

The problem with the above solution is that it does not satisfy the constraints imposed by 
combining Eq.~(\ref{Eq:constraint2}) with Eq.~(\ref{Eq:5othertransmissions}).
Specifically, it does not satisfy the constraints 
\begin{subequations}
\label{Eq:important-constraints}
\begin{eqnarray}
0 &\leq&{\cal T}_{\rm LM}(\eps) +{\cal T}_{\rm LR}(\eps) -{\cal T}_{\rm RL}(\eps)  \ \ \leq\ \  N^{\rm min}_{\rm ML},
\label{Eq:important-constraints-a}
\\
0 &\leq& {\cal T}_{\rm RM}(\eps)+{\cal T}_{\rm RL}(\eps)-{\cal T}_{\rm LR}(\eps) \ \ \leq\ \ N^{\rm min}_{MR}.
\label{Eq:important-constraints-b}
\end{eqnarray}
\end{subequations}
The proposed solution violates the lower bound in Eq.~(\ref{Eq:important-constraints-a}) for all $\eps > \eps_{\rm 1R}$.  
Similarly, it violates the lower bound in Eq.~(\ref{Eq:important-constraints-b}) for all $\eps < \eps_{\rm 1L}$.  
In addition, in the case of a refrigerator with $\eps_{\rm 0R} > \eps_{\rm 1R}$, as in Fig.~\ref{Fig:boxcars}b, then the 
proposed solution violates the upper bound in Eq.~(\ref{Eq:important-constraints-b}) for all $\eps_{\rm 1R} < \eps < \eps_{\rm 0R}$. Similarly, when  $\eps_{\rm 1L} > \eps_{\rm 0L}$, the solution also violates the upper bound in Eq.~(\ref{Eq:important-constraints-a}) for all $\eps_{\rm 0L} < \eps < \eps_{\rm 1L}$. 
We will fix this in the case of a heat-engine by explicitly adding these bounds in the next section, the case of
a refrigerator will be treated in Section~\ref{Sect:fri}.

\subsection{Optimizing transmissions while respecting all constraints}
\label{Sect:constraints}

Here, we consider carrying out the optimization given by the list (a-d) in the previous section
within the limits given by the constraints in Eq.~(\ref{Eq:important-constraints}).
As we are considering a heat-engine, we know that the $\eps_{\rm 0L}$, $\eps_{\rm 1L}$, $\eps_{\rm 0R}$ and $\eps_{\rm 1R}$ are ordered as in Eq.~(\ref{Eq:eng-eps-ordering}), see Fig.~\ref{Fig:boxcars}a. 
The optimization for $\eps$ in the window between  $\eps_{\rm 1L}$ and $\eps_{\rm 1R}$ is trivial, since 
there the transmission functions in the above list (a-d) does not violate the constraints in Eq.~(\ref{Eq:constraint2}).
This leaves us with the less trivial part of the optimization under the constraints,
for $\eps > \eps_{\rm 1R}$ and $\eps<\eps_{\rm 1L}$.

 \subsubsection{{\it Optimization for $\eps > \eps_{\rm 1R}$ or $\eps<\eps_{\rm 1L}$}}
 \label{Sect:optimization+constraint-large-eps}
 
 For $\eps > \eps_{\rm 1R}$ the independent optimization of the transmission functions, required increasing ${\cal T}_{\rm RL}$ while decreasing 
 ${\cal T}_{\rm LM}$ and ${\cal T}_{\rm LR}$ but doing this comes into conflict with the constraint that 
 ${\cal T}_{\rm LR} \geq {\cal T}_{\rm RL} - {\cal T}_{\rm LM}$ due to Eq.~(\ref{Eq:important-constraints-a}).
 Thus we do the unconstrained optimization in the previous section up to the point allowed by the constraint,
 after which
 \begin{eqnarray}
 {\cal T}_{\rm LR}(\eps) ={\cal T}_{\rm RL}(\eps)  - {\cal T}_{\rm LM}(\eps) \, .
 \end{eqnarray}
 We then ask if $J_{\rm M}$ decreases (for fixed power generation) when we increase slice $\gamma$ of 
 ${\cal T}_{\rm RL}$ and ${\cal T}_{\rm LM}$ by infinitesimal amounts $\delta\tau^{(\gamma)}_{\rm RL}$ and $\delta\tau^{(\gamma)}_{\rm LM}$ respectively, given that one must also change slice $\gamma$ of 
 ${\cal T}_{\rm LR}$ by $\delta\tau^{(\gamma)}_{\rm LR}=\delta\tau^{(\gamma)}_{\rm RL}-\delta\tau^{(\gamma)}_{\rm LM}$ not to violate the above constraint.
With this observation, we find that for $J_{\rm M}$ to decrease we need
\begin{eqnarray}
& &\delta\tau^{(\gamma)}_{\rm RL} \ (\eps_{\rm 1R} -\eps_{\rm 1L})\, \left[ f_{\rm R}(\eps_\gamma)- f_{\rm L}(\eps_\gamma) \right] 
\ +\ \delta\tau^{(\gamma)}_{\rm LM}\ (\eps_\gamma-\eps_{\rm 1L})\,\left[ f_{\rm M}(\eps_\gamma)- f_{\rm R}(\eps_\gamma) \right] 
\ <\ 0.
\end{eqnarray}
Since all the brackets in the above expression are positive for $\eps_\gamma > \eps_{\rm 0R}$, we see that to minimize
$J_{\rm M}$ we should minimize both ${\cal T}_{\rm RL}$ and ${\cal T}_{\rm LM}$.
Thus we conclude that for $\eps > \eps_{\rm 1R}$, it is optimal that all transmission functions are zero.

The situation where $\eps < \eps_{\rm 1L}$ can be treated in the same manner as above, 
upon interchanging the labels ``L'' and ``R''.
Thus, the optimal situation is when all transmission functions are zero for $\eps < \eps_{\rm 1L}$.

\subsubsection{{\it Conclusion of optimization with constraints}}
\label{Eq:conclusion-of-optimization}

Bringing together the results found so far, the transmission functions which maximize the heat-engine's efficiency 
for a given power generation are
\begin{subequations}
\label{Eq:final-Ts}
\begin{eqnarray}
{\cal T}_{\rm RL}(\eps) \ =\ {\cal T}_{\rm LR}(\eps) \ = \ 0  \ \qquad \hbox{ for all } \eps ,
\qquad \qquad \quad & & 
\label{Eq:final-T_RL}\\
{\cal T}_{\rm RM}(\eps) \ =\ 
\left\{ \begin{array}{ccl}  N^{\rm min}_{\rm RM}
& & \hbox{ for }\ \eps_{\rm 0R} \leq \eps \leq \eps_{\rm 1R}, \\
0 & & \hbox{ otherwise,} 
\end{array} \right. & &
\\
{\cal T}_{\rm LM}(\eps) \ =\  
\left\{ \begin{array}{ccl} 
 N^{\rm min}_{\rm LM} &  & \hbox{ for }\ \eps_{\rm 1L} \leq \eps \leq \eps_{\rm 0L}, \\
0 & & \hbox{ otherwise,} 
\end{array} \right. & & \quad
\end{eqnarray}
where $N^{\rm min}_{ij}$ is defined in Eq.\ (\ref{Eq:Nmin}).
These functions are sketched in Fig.~\ref{Fig:boxcars}c.
Eq.~(\ref{Eq:final-T_RL}) means the optimal system has no direct flow of electrons between reservoirs L and R. 
Given Eqs.~(\ref{Eq:T_ML},\ref{Eq:T_MR}), this means that
\begin{eqnarray}
{\cal T}_{\rm MR}(\eps) ={\cal T}_{\rm RM}(\eps)
\ \hbox{ \& } \
{\cal T}_{\rm ML}(\eps) = {\cal T}_{\rm LM}(\eps)  \ \hbox{ for all } \eps . \quad 
\end{eqnarray}
\end{subequations}
Hence, the optimal  three terminal situation is one that can be thought of as a pair of two-terminal problems
much like those already  considered in 
Refs.~\cite{whitney-prl2014,whitney2015}. 
To be more explicit, Eqs.~(\ref{Eq:final-Ts}) tell us that  the optimal transmission is one that can be split into 
a problem of optimizing transmission between M and R through $N^{\rm min}_{\rm RM}$ transverse modes (with ${\cal T}_{\rm MR}(\eps) ={\cal T}_{\rm RM}(\eps)$ at all $\eps$)
and another problem of optimizing transmission between M and L through $N^{\rm min}_{\rm LM}$ transverse modes
 (with ${\cal T}_{\rm ML}(\eps) ={\cal T}_{\rm LM}(\eps)$ at all $\eps$).
These two optimization problems could be treated independently were it not for the fact they are coupled
by the constraint that  the electrical currents in the two problems $I_{\rm L} $ and $I_{\rm R}$ must sum to zero
to get Eq.~(\ref{Eq:IM=0}).

\section{Showing the three-terminal system cannot exceed the bound for two-terminal systems}
 \label{Sect:over-estimate}

 Given the previous section's observation that the optimal transmission for a three-terminal
 system is one which can be split into a pair of two-terminal transmission problems, 
 we can draw two conclusions.

 Firstly, the optimal transmission for a three-terminal system does not require any time-reversal symmetry breaking of the type generated by an external magnetic field.  Thus, the optimal transmission can be achieved in a system without an external magnetic field.
We wish to be clear that this proof does not mean that magnetic fields may not be helpful in specific situations; for example, a magnetic field may be helpful in tuning the transmission of a given system to be closer to the optimal one.
However, it does mean that there is no requirement to have a magnetic field; other parameters (which do not break time-reversal symmetry) can be tuned to bring the system's transmission to the optimal one.
This is the first main conclusion of this work.

Secondly, it is not hard to show that
 a three-terminal system cannot exceed the bounds found in 
Refs.~\cite{whitney-prl2014,whitney2015} for a pair of two-terminal systems
with the same number of transverse modes.
To be more specific, it cannot exceed the bound for a pair of two-terminal systems where one of the two-terminal systems has $N^{\rm min}_{\rm LM}$ transverse modes and the
other has $N^{\rm min}_{\rm RM}$ transverse modes, see Eq.~(\ref{Eq:Nmin}).
To prove this bound,
it is sufficient to remark that the optimization of the three-terminal system in Eq.~(\ref{Eq:final-Ts}) is exactly that of the optimization of a pair of two-terminal systems, with an additional constraint
that the electrical currents in the two problems ($I_{\rm L} $ and $I_{\rm R}$) sum to zero.  
This constraint couples the two problems and makes them much harder to resolve.  However, if we simply drop the constraint
on $I_{\rm L}$ and $I_{\rm R}$ and perform the optimization, we can be certain that we are over-estimating the efficiency at given power output.  Once we drop this constraint the two optimization problems become completely decoupled from each other. Thus, we can optimize the transmission between M and R using the method in Refs.~\cite{whitney-prl2014,whitney2015}, and independently optimize the transmission between M and L
using the same method.  As a result, an over-estimate of the three-terminal efficiency at given power output is bounded by the maximum two-terminal efficiency of a pair of two-terminal systems, with this bound being the one found in 
Refs.~\cite{whitney-prl2014,whitney2015}.
This is the second main conclusion of this work.

\section{Achieving the two-terminal bound in a three-terminal system}
\label{Sect:achieving}

Having found an upper bound on the efficiency at given power output by using a process that over-estimates the efficiency, we can be sure that no three terminal system can be {\it more} efficient than a pair of optimal two-terminal systems.  This makes it  natural to ask if any three terminal system can be {\it as} efficient as this pair of optimal two-terminal systems.
To answer this question, we present an example of a three-terminal system which is as efficient as the pair of optimal two-terminal systems.  This will be our proof that the upper bound on the efficiency of a three-terminal system coincides with the upper bound on the efficiency of a pair of two-terminal systems.

To proceed we take a three-terminal system with $N^{\rm min}_{\rm LM}=N^{\rm min}_{\rm RM}$.
Given Eq.~(\ref{Eq:Nmin}), this could be a system with $N_{\rm L}=N_{\rm R}$,
or it could be a system with $N_{\rm M}$ less than both $N_{\rm L}$ and $N_{\rm R}$,
In this case, one can take a pair of optimal two-terminal solutions from Refs.~\cite{whitney-prl2014,whitney2015}, 
in the cases where $\eminus V_{\rm R}=-\eminus V_{\rm L} >0$.
They have
\begin{subequations}
\label{Eq:eps-2term}
\begin{eqnarray}
\eps_{\rm 0L}=-\eps_{\rm 0R} \  \hbox{ \& }  \ \eps_{\rm 1L}=-\eps_{\rm 1R}\ ,
\end{eqnarray}
with 
\begin{eqnarray}
\eps_{\rm 0R} = {\eminus V_R \over  1- T_{\rm R}/ T_{\rm M} }
\ \hbox{ \& }\ 
\eps_{\rm 1R}= \eminus V_R {\partial_{\rm R} J_M^{\rm (R)} \over \partial_{\rm R} P^{\rm (R)}_{\rm gen}}  \ ,
\label{Eq:eps1-two-term}
\end{eqnarray}
\end{subequations}
where we have written the results of Refs.~\cite{whitney-prl2014,whitney2015}
in terms of the notation of this article, with the derivatives defined in Eq.~(\ref{Eq:define-dL-and-dR}).
Here, $ J_M^{(i)}$ is the part of the heat carried out of reservoir M by electron flow between reservoir M and 
reservoir $i$,  and   $P^{(i)}_{\rm gen}$ is the part of the total power generated
by that electron flow, so
\begin{subequations}
\begin{eqnarray}
J_{\rm M}&=& J^{\rm (R)}_{\rm M} + J^{\rm (L)}_{\rm M} \, ,
\label{Eq:J-parts}
\\
P_{\rm gen} &=& P^{\rm (R)}_{\rm gen}+P^{\rm (L)}_{\rm gen} \, .
\end{eqnarray}
\end{subequations}
Conservation of electrical current gives $I_{\rm M} = -I_{\rm L}-I_{\rm R}$.
As the only dependence on $V_i$ within $I_{\rm M}$, $J_{\rm M}$ and $P_{\rm gen}$ 
are in $I_i$, $J^{(i)}_{\rm M}$ and $P^{(i)}_{\rm gen}$, respectively, we have
\begin{eqnarray}
\partial_{\rm i} I_{\rm M} = -\partial_{\rm i} I_{\rm i}\, , \  
\partial_{\rm i} J_{\rm M} = \partial_{\rm i} J^{(i)}_{\rm M} \  \hbox{ \& }  \ 
\partial_{\rm i} P_{\rm gen} = \partial_{\rm i} P^{(i)}_{\rm gen}\, . \quad
\end{eqnarray}
With some thought about the symmetries between L and R, we see that 
the derivatives have the following symmetries between L and R,
\begin{subequations}
\label{Eq:symmetries-of-partials}
\begin{eqnarray}
\partial_{\rm L} I_{\rm M} &=&  \partial_{\rm R} I_{\rm M}\ ,
\\
\partial_{\rm L} J_{\rm M} &=&  -\partial_{\rm R} J_{\rm M}\ ,
\\
\partial_{\rm L} P_{\rm gen} &=&  -\partial_{\rm R} P_{\rm gen}\ .
\end{eqnarray}
\end{subequations}

We recall that Eqs.~(\ref{Eq:eps-2term}-\ref{Eq:symmetries-of-partials}) are all for an optimal pair of 
{\it two-terminal} systems. 
We now take the information in  Eqs.~(\ref{Eq:eps-2term}-\ref{Eq:symmetries-of-partials}), and verify that they {\it also} give an optimal solution of the three-terminal problem.
For this we note that the definition of $\eps_{\rm 0R}$ and $\eps_{\rm 0L}$ are the same in the two- and three-terminal problems, however the definition of $\eps_{\rm 1R}$ and $\eps_{\rm 1L}$ are different, with 
that for three-terminals being Eq.~(\ref{Eq:eps1}) and that for two-terminals
being Eq.~(\ref{Eq:eps1-two-term}).
However, if we now take the symmetry relations in Eq.~(\ref{Eq:symmetries-of-partials}), we see that 
 Eq.~(\ref{Eq:eps1})  reduces to  Eq.~(\ref{Eq:eps1-two-term}). 
Thus, the solution of the optimization problem for a pair of two-terminal systems in Eqs.~(\ref{Eq:eps-2term}-\ref{Eq:symmetries-of-partials}),  is {\it also} a solution of the optimization problem
for the three-terminal problem.  All currents are the same in the three-terminal system as in the pair of two-terminal systems, so the efficiency and power output are also the same.
Finally, we note that this solution has $I_{\rm L}=-I_{\rm R}$, so it satisfies $I_{\rm M}=0$ as in Eq.~(\ref{Eq:IM=0}). 
Hence, we have shown that an optimal three-terminal system can be as good as a pair of optimal two-terminal systems.  This is the third main conclusion of this work (after the two in the previous section).

Combining this conclusion  with the others, we find that the upper bound on efficiency
at given power output is the same for a three-terminal system as for a pair of two terminal systems.
This means that  the optimal three-terminal system has no advantage over a pair
of optimal two-terminal systems,  however it does not tell us in which geometry it is easier to engineer a system 
achieves (or gets close to) that optimum.

\section{Route to the optimal transmission for $N^{\rm min}_{\rm LM}\neq N^{\rm min}_{\rm RM}$}
\label{Sect:Optimization-procedure}

We can use the results of the two preceding sections to get a simple over-estimate of the
maximal efficiency at given power generation for  a machine with $N^{\rm min}_{\rm LM}\neq N^{\rm min}_{\rm RM}$.  This upper bound is given by the efficiency of an {\it equivalent}  three-terminal machine with $N^{\rm min}_{\rm LM}= N^{\rm min}_{\rm RM}$.   Here, we define an ``equivalent'' system as one with the same $N^{\rm min}_{\rm LM}+ N^{\rm min}_{\rm RM}$.  
In the case where  $N_{\rm M} > N_{\rm L},N_{\rm R}$, this is the same as saying that for given $N_{\rm L}+N_{\rm R}$ an optimal machine with $N_{\rm L}\neq N_{\rm R}$ cannot
be better than an optimal machine with $N_{\rm L}=N_{\rm R}$. While for $N_{\rm M} < N_{\rm L},N_{\rm R}$,
all systems have $N^{\rm min}_{\rm LM}= N^{\rm min}_{\rm RM}$.
However, it is likely that this upper bound for $N^{\rm min}_{\rm LM}\neq N^{\rm min}_{\rm RM}$ 
is clearly an over-estimate, since it is probably only for 
$N^{\rm min}_{\rm LM}= N^{\rm min}_{\rm RM}$ that the 
optimal efficiency with the constraint that $I_{\rm M}=0$ is as large as that without this constraint.
This greatly reduces practical interest in optimizing a system with $N^{\rm min}_{\rm LM}\neq N^{\rm min}_{\rm RM}$, since optimizing implies a significant amount of control over the system, in which case it is better to
engineer the system to have $N^{\rm min}_{\rm LM}= N^{\rm min}_{\rm RM}$, and optimize that.

If we wished, we could get a strict upper-bound on efficiency at given power generation for
a system with given $N^{\rm min}_{\rm LM}\neq N^{\rm min}_{\rm RM}$.
However, the optimization procedure for this is heavy, 
as well of being of little practical interest. Thus, we do not carry it out here,
we simply list the principle steps. 
\begin{itemize}
\item[(i)] 
Write explicit results for the currents and power in terms of four parameters $\eps_{\rm 1L}$, $\eps_{\rm 1R}$, $V_{\rm L}$ and $V_{\rm R}$ (noting that $\eps_{\rm 0L}$ and $\eps_{\rm 0R}$ are given by $V_{\rm L}$ and $V_{\rm R}$ in Eq.~(\ref{Eq:eps0})). 
Use these to calculate the derivatives that appear on the right hand side of Eq.~(\ref{Eq:eps1}), getting them as explicit functions of $\eps_{\rm 1L}$, $\eps_{\rm 1R}$, $V_{\rm L}$ and $V_{\rm LR}$.  This step is straight-forward, and is carried out in Appendix~\ref{Sect:currents-powers-derivatives}.

\item[(ii)]
Substitute these derivatives into the right hand side of Eq.~(\ref{Eq:eps1})
for $i={\rm L}$ and $i={\rm R}$, this gives a pair of transcendental equations 
for the four parameters $\eps_{\rm 1L}$, $\eps_{\rm 1R}$, $V_{\rm L}$ and $V_{\rm R}$.
Since we are interested in $I_{\rm L}=-I_{\rm R}$, with $I_{\rm L}$ and $I_{\rm R}$ being algebraic functions calculated in step (i) above (see Appendix~\ref{Sect:currents-powers-derivatives}), this gives a 
third transcendental equation for these four parameters.
 
 \item[(iii)]
Solve the three simultaneous transcendental equations numerically.
As we have four unknown parameters and only three equations, we will get three parameters in terms of the fourth.
We propose getting  $\eps_{\rm 1L}$, $\eps_{\rm 1R}$, and $V_{\rm L}$ as functions of  $V_{\rm R}$.  This involves solving the set of three simultaneous equations once for each value of $V_{\rm R}$. This is the heavy part of the calculation, which one would have to perform numerically. We do not do this here. 

\item[(iv)]
Once we have $\eps_{\rm 1L}$, $\eps_{\rm 1R}$, and $V_{\rm L}$ as a function of $V_{\rm R}$,
we can get all electrical and heat currents as a function of $V_{\rm R}$ alone. 
Since step (iii) was performed numerically, we are forced to do this step numerically as well.
The electrical currents give us the power generated, $P_{\rm gen}$, as a function of the voltage $V_{\rm R}$,
which we must invert (again numerically) to get the voltage as a function of the power generated,   
$V_{\rm R}(P_{\rm gen})$.
We then take the result for $J_{\rm M}$ as a function of $V_{\rm R}$, and substitute in $V_{\rm R}(P_{\rm gen})$.
This will give us $J_{\rm M}(P_{\rm gen})$, the optimal (minimum) heat flow out of reservoir M for a given power generated.
Then the maximal heat-engine efficiency 
$\eta_{\rm eng}(P_{\rm gen}) = P_{\rm gen} \big/ J_{\rm M}(P_{\rm gen})$.

\end{itemize}

\section{Maximum refrigerator efficiency for given cooling power}
\label{Sect:fri}

In Refs.~\cite{whitney-prl2014,whitney2015} an upper bound on refrigerator efficiency for {\it given cooling power} was calculated directly for two-terminal devices.  The result looked extremely similar to those works' result for the upper bound on heat-engine efficiency for {\it given power output}.  It has since become clear to us how to 
get the result for refrigerators from the result for heat-engines.  The trick it to make the physically plausible assumption that the upper bound on the cooling power  of a refrigerator, $J_{\rm M}$, is a monotonic function of the electrical power it absorbs,  $P_{\rm abs}$.
Then the curve of maximum efficiency versus cooling power, $J_{\rm M}$, is the same as the curve of
maximum efficiency versus absorbed power $P_{\rm abs}$ (upon transforming the horizontal axis from $P_{\rm abs}$ to $J_{\rm M}$ using the maximal efficiency curve).
This is a great simplification of the problem, as it turns out that 
finding the refrigerator with maximal efficiency at given absorbed power, is a rather straightforward extension of the above calculation of the optimal heat-engine at given power output.

Here we take this point of view, we find the three-terminal refrigerator with maximal efficiency for given absorbed power, by a few straightforward modifications of the heat-engine calculation.
A system absorbing power $P_{\rm abs}$ is the same as a system generating negative power $P_{\rm gen}=-P_{\rm abs}$.  The crucial modification is that we must {\it maximize} $J_{\rm M}$ at given negative $P_{\rm gen}$ for refrigerators, when we were {\it minimizing}
 $J_{\rm M}$ at given positive $P_{\rm gen}$  for heat-engines.

Inspecting the calculation in Section~\ref{Sect:eng}, we see that everything follows through 
for a refrigerator with $T_{\rm L}=T_{\rm R}=T_0$, $T_{\rm M} < T_0$, and  $\eminus V_{\rm L} \, > \, 0\, >\, \eminus V_{\rm R}$. Except that now
we maximize $J_{\rm M}$, and that now 
the Fermi functions in this case are those sketched in Fig.~\ref{Fig:fermi-functions}b, obeying
\begin{eqnarray}
\big[f_{\rm R}(\eps)-f_{\rm L}(\eps)\big]& \hbox{ is} &  \hbox{negative for all } \eps \ ,
\label{Eq:fri-sign-of-F-difference2}
\\
\big[f_{\rm M}(\eps)-f_i(\eps)\big] & \hbox{ is} &
\left\{ \begin{array}{l}
\hbox{ negative for $\eps >\eps_{0i}$} \,,\\
\hbox{ positive for $\eps <\eps_{0i}$} \,,
\end{array}\right.
\label{Eq:fri-sign-of-F-difference1}
\end{eqnarray}
where Eq.~(\ref{Eq:eps0}) is more conveniently written as
\begin{eqnarray}
\eps_{0i} = {-\eminus V_i \over T_0/T_{\rm M} -1}\ .
\label{Eq:eps0-fri}
\end{eqnarray}

By a careful comparison with Section~\ref{Sect:eng}, we note that all relevant differences of Fermi functions in the refrigerator case have the opposite sign from in the heat-engine case.  Thus, if a given change of transmission reduces $J_{\rm M}$ for the heat-engine, then that same change will increase $J_{\rm M}$ for the refrigerator.
Hence, we conclude that the procedure that optimizes a heat-engine (minimizing $J_{\rm M}$ for given $P_{\rm gen}$ and $I_{\rm M}$) also optimizes a refrigerator (maximizing $J_{\rm M}$ for given $P_{\rm gen}$ and $I_{\rm M}$).

The independent optimization of ${\cal T}_{\rm RM}$,  ${\cal T}_{\rm LM}$,  ${\cal T}_{\rm RL}$ 
and ${\cal T}_{\rm LR}$ follows exactly as in Section~\ref{Sect:independent}.
As with the heat-engine, it is difficult to guess the values of $\eps_{\rm 1R}$ and $\eps_{\rm 1L}$
from their definition in Eq.~(\ref{Eq:eps1}).
However, for maximal refrigeration we want both terms in $P_{\rm gen}$ in Eq.~(\ref{Eq:Pgen-integral})
to be negative (so the absorbed power $P_{\rm abs}=-P_{\rm gen}>0$). By inspection of Eqs.~(\ref{Eq:I_L-integral},\ref{Eq:I_R-integral}) we see that this requires $\eps_{\rm 1R}< \eps_{\rm 0R}$ and $\eps_{\rm 1L}>\eps_{\rm 0L}$.
Further, we can see that $\eps_{\rm 1L}$ must be negative. To do this we inspect the terms in 
Eqs.~(\ref{Eq:J_M-integral},\ref{Eq:Pgen-integral}) 
which depend on $\eps_{\rm 1L}$, and we see that making $\eps_{\rm 1L}$ positive will increase $P_{\rm abs}$, while reducing the cooling power $J_{\rm M}$, which is clearly not a way to maximize the efficiency, $\eta_{\rm fri}$.
A similar argument convinces us that  $\eps_{\rm 1R}$ must be positive.
Thus, we are interested in the case summarized in Eq.~(\ref{Eq:fri-eps-ordering}).

\subsection{Optimizing refrigerator while respecting all constraints}
\label{Sect:fri-constraints}
 
 As we have $\eps_{\rm 0L} < \eps_{\rm 1L} < 0 < \eps_{\rm 1R} < \eps_{\rm 0R}$,
 the result of independently optimizing the transmission functions is that shown in Fig.~\ref{Fig:boxcars}b.
 For $\eps$ between $\eps_{\rm 1L}$ and $\eps_{\rm 1R}$, no constraint are violated by that result; so the optimal solution remains that all
 transmission functions are zero in this window.
 The optimization for $\eps >\eps_{\rm 0R}$ and $\eps < \eps_{\rm 0L}$ follows the same logic as in Section~\ref{Sect:optimization+constraint-large-eps}, except that now we want to maximize $J_{\rm M}$ and the differences of Fermi functions have the opposite signs.   We find that the system is optimized by having all  transmission  functions equal to zero for $\eps > \eps_{\rm 0R}$ and for $\eps < \eps_{\rm 0L}$.

 \subsubsection{{\it Optimization for $\eps$ between $\eps_{\rm 0R} $ and $\eps_{\rm 1R}$.}}
\label{Sect:optimization+constraint-smaller-eps}
 
For $\eps$ in the window $\eps_{\rm 1R} < \eps < \eps_{\rm 0R}$, the 
independent  optimization (maximizing ${\cal T}_{\rm RM}$ and ${\cal T}_{\rm RL}$, while minimizing all other transmissions) violates both the lower bound in Eq.~(\ref{Eq:important-constraints-a}) and the upper bound in Eq.~(\ref{Eq:important-constraints-b}).
This case must be treated with care.
We start by increasing ${\cal T}_{\rm RM}$ and ${\cal T}_{\rm RL}$ while reducing ${\cal T}_{\rm LM}$ and ${\cal T}_{\rm LR}$, until we reach the limit of the bounds in Eqs.~(\ref{Eq:important-constraints-a}) and (\ref{Eq:important-constraints-b}); this occurs at
\begin{eqnarray}
{\cal T}_{\rm LM}(\eps) &=& {\cal T}_{\rm RL}(\eps) - {\cal T}_{\rm LR}(\eps),
\\
{\cal T}_{\rm RM}(\eps) &=& -{\cal T}_{\rm RL}(\eps) + {\cal T}_{\rm LR}(\eps) +N^{\rm min}_{\rm MR}(\eps).
\end{eqnarray} 
 We then ask if $J_{\rm M}$ increases (for fixed $P_{\rm abs}$) when we increase slice $\gamma$ of 
 ${\cal T}_{\rm RL}$ and ${\cal T}_{\rm LR}$ by infinitesimal amounts $\delta\tau^{(\gamma)}_{\rm RL}$ and $\delta\tau^{(\gamma)}_{\rm LR}$ respectively, given that the above constraint means that one must also change slice $\gamma$ of 
 ${\cal T}_{\rm LM}$ by $\delta\tau^{(\gamma)}_{\rm LM}=\delta\tau^{(\gamma)}_{\rm RL}-\delta\tau^{(\gamma)}_{\rm LM}$, and change slice $\gamma$ of 
 ${\cal T}_{\rm RM}$ by $\delta\tau^{(\gamma)}_{\rm LM}=-\delta\tau^{(\gamma)}_{\rm RL}+\delta\tau^{(\gamma)}_{\rm LM}$.
With this observation, we find that for $J_{\rm M}$ to increase we need  
\begin{eqnarray}
& &\delta\tau^{(\gamma)}_{\rm RL} \ (\eps_{\rm 1R} -\eps_{\rm 1L})\, \left[ f_{\rm M}(\eps_\gamma)- f_{\rm L}(\eps_\gamma) \right] 
\ +\  
\delta\tau^{(\gamma)}_{\rm LR}\ (\eps_{\rm 1R}-\eps_{\rm 1L})\,\left[ f_{\rm R}(\eps_\gamma)- f_{\rm M}(\eps_\gamma) \right] 
\ >\ 0.
\end{eqnarray}
Since all brackets in the above expression are negative  for $\eps_\gamma < \eps_{\rm 0R}$, we see that to maximize
$J_{\rm M}$ we should minimize both ${\cal T}_{\rm RL}$ and ${\cal T}_{\rm LR}$.
Thus the optimum for $\eps$ between $\eps_{\rm 1R}$ and $\eps_{\rm 0R}$ is that ${\cal T}_{\rm RM}$ is maximal
(${\cal T}_{\rm RM} = N^{\rm min}_{\rm RM}$) while the other transmission functions are zero.

The same logic can be applied to the energies $\eps$ between  $\eps_{\rm 0L}$ and $\eps_{\rm 1L}$,
and we conclude that the optimal there is that ${\cal T}_{\rm LM}$ is maximal
(${\cal T}_{\rm LM} = N^{\rm min}_{\rm LM}$) while the other transmission functions are zero.

\subsubsection{{\it Conclusion of optimization with constraints}}
\label{Eq:fri-conclusion-of-optimization}

To summarize, the transmission functions
which maximize refrigerator cooling power $J_{\rm M}$ for given absorbed power $P_{\rm abs}$
are
\begin{subequations}
\label{Eq:fri-final-Ts}
\begin{eqnarray}
{\cal T}_{\rm RL}(\eps) \ =\ {\cal T}_{\rm LR}(\eps) \ = \ 0  \ \qquad \hbox{ for all } \eps ,
\qquad \qquad \quad & & 
\\
{\cal T}_{\rm RM}(\eps) \ =\ 
\left\{ \begin{array}{ccl}  N^{\rm min}_{\rm RM}
& & \hbox{ for }\ \eps_{\rm 1R} \leq \eps \leq \eps_{\rm 0R}, \\
0 & & \hbox{ otherwise,} 
\end{array} \right. & &
\\
{\cal T}_{\rm LM}(\eps) \ =\  
\left\{ \begin{array}{ccl} 
 N^{\rm min}_{\rm LM} &  & \hbox{ for }\ \eps_{\rm 0L} \leq \eps \leq \eps_{\rm 1L}, \\
0 & & \hbox{ otherwise,} 
\end{array} \right. & & \quad
\end{eqnarray}
where $N^{\rm min}_{ij}$ is defined in Eq.~(\ref{Eq:Nmin}).
These transmission functions are sketched in Fig.~\ref{Fig:boxcars}d.
Given these results and Eqs.~(\ref{Eq:T_ML},\ref{Eq:T_MR})
we also have 
\begin{eqnarray}
{\cal T}_{\rm MR}(\eps) ={\cal T}_{\rm RM}(\eps)
\ \hbox{ \& }\ 
{\cal T}_{\rm ML}(\eps) = {\cal T}_{\rm LM}(\eps)  \hbox{ for all } \eps . \quad 
\end{eqnarray}
\end{subequations}

Every statement made in Sections~\ref{Sect:over-estimate} and \ref{Sect:achieving} about heat-engines has its analogue for refrigerators.
In particular, we have proven that direct transmission between left and right is detrimental to the efficiency of the refrigerator.
Once this left-right transmission is suppressed, the  three terminal problem for a refrigerator can be thought of as a pair of two-terminal problems of the form in Refs.~\cite{whitney-prl2014,whitney2015}. 
The role of chirality is then irrelevant in the refrigerator, by which we mean that the optimal transmission can be achieved with or without the time-reversal symmetry breaking that an external magnetic field induces. 
We can use exactly the same logic as applied to the heat-engine in Section~\ref{Sect:over-estimate} 
to say that a three-terminal refrigerator cannot exceed the upper bound on efficiency for given cooling power given
in Refs.~\cite{whitney-prl2014,whitney2015}, for a pair of two-terminal thermoelectric refrigerators (one with $N^{\rm min}_{\rm LM}$ transverse modes and the other with 
$N^{\rm min}_{\rm RM}$ transverse modes).
As in Section~\ref{Sect:achieving}, this two-terminal bound can be achieved in a three-terminal refrigerator with  
$N^{\rm min}_{\rm LM}=N^{\rm min}_{\rm RM}$.

\section{Minimal entropy production for given power output}

Ref.~\cite{whitney2015,Cleuren2012} showed that the efficiency at given power immediately gives the entropy production at that power.  The rate of entropy production of a heat-engine at power output,
$P_{\rm gen}$, is
\begin{eqnarray}
\dot S (P_{\rm gen}) 
 &=& 
{P_{\rm gen} \over T_R} \left({\eta_{\rm eng}^{\rm Carnot} \over \eta_{\rm eng}(P_{\rm gen})} -1 \right),
\label{Eq:dotS-eng}
\end{eqnarray}
where $\eta_{\rm eng}^{\rm Carnot}$ is given in Eq.~(\ref{Eq:Carnot-eng}).
While for a refriegrator
at cooling power $J_L$, it is 
\begin{eqnarray}
\dot S (J_L) 
 &=& 
{J_L \over T_R} \left({1\over \eta_{\rm fri}(J_L)} -{1 \over \eta_{\rm fri}^{\rm carnot}}  \right),
\label{Eq:dotS-fri}
\end{eqnarray}
where $\eta_{\rm fri}^{\rm carnot}$ is given in Eq.~(\ref{Eq:Carnot-fri}).
It is straight-forward to prove that these formulas apply equally to the three-terminal systems that we consider here.
Hence, an upper bound on efficiency at given power output immediately gives a lower bound on the rate of entropy production at that power output.  This means that the results in this work also tell us that the lower bound on entropy production by a three-terminal system at given power output is the same as the lower bound on two-terminal systems
discussed in Ref.~\cite{whitney2015}.

\section{Concluding remarks}

We have used scattering theory to find the upper bound on the efficiency of a three-terminal thermoelectric quantum machine at given power output.  We find that this bound can be achieved at any external magnetic fields,
so the bound is the same for chiral thermoelectrics as for those with no external field.
This upper bound on efficiency is identical to that found for two-terminal thermoelectric systems in 
Refs.~\cite{whitney-prl2014,whitney2015}. 
It equals the Carnot efficiency when the power output is 
zero, but it decays monotonically for increasing power output, as shown in
Fig.~2 of Ref.~\cite{whitney2015}.

We wonder if one can derive the similar bound for the system in Fig.~\ref{Fig:three-term}c 
with a microscopic model of the photon (or phonon) exchange, rather than the phenomenological 
model used here.

Most real quantum systems also lose heat to the environment (through photon or phonon exchange), 
this can be modelled as a fourth terminal
which exchanges heat but not charge with the system.   A similar four-terminal geometry was discussed in Ref.~\cite{wshs2016}, which showed that such system operates in a non-thermal state and so exhibits non-local laws of thermodynamics.  
It would be interesting to see how this bound behaves in such a situation,
although we doubt that the pedestrian (brute-force) optimization used in this work will be extendable to more than three-terminals.

\appendix

\vskip 3mm 
\section{Currents, powers and their derivatives in terms of $\eps_{0i}$ and $\eps_{1i}$}
\label{Sect:currents-powers-derivatives}

In what follows, it is useful to define two functions,
\begin{eqnarray}
G_j(\eps) \equiv \int_\eps^\infty { \rmd \tilde\eps \over h} \ f_j(\tilde\eps),  \qquad \qquad
F_j(\eps) \equiv \int_\eps^\infty { \rmd \tilde\eps \over h}\ \tilde\eps \ f_j(\tilde\eps).
\label{Eq:G_j+F_j}
\end{eqnarray}
The first of these integrals can be evaluated by defining $x_j=(\eps-\eminus V_j) \big/  (\kB T_j)$, 
so
 \begin{eqnarray}
G_j(\eps)
= {\kB T_j \over h}\int_{x_j}^{\infty} {\rmd x \ \e^{-x} \over 1+\e^{-x}}
=  {\kB T_j \over h} \ln \left[1+\e^{-x_j}\right]. 
\label{Eq:G_j-result}
\end{eqnarray}
With a shift of integration variable, we find that
 \begin{eqnarray}
 F_j(\eps)
&=& {(\kB T_j )^2\over h}
\int_0^{\infty} {\rmd x \ \big(x+\eps_0/(\kB T_j)\big) \over 1+\e^{x + x_j}} 
\ =\ 
\eps G_j(\eps)  -{(\kB T_j)^2 \over h} {\rm Li}_2\left(-\e^{-x_j}\right) ,
\label{Eq:F_j-result}
\end{eqnarray}
where the dilogarithm function ${\rm Li}_2(t) = \int_0^\infty \rmd x \, x \, (\e^x/t -1)^{-1}$.

Eqs.~(\ref{Eq:I_L-integral}-\ref{Eq:I_H-integral}) with
Eqs.~(\ref{Eq:final-Ts}) give
\begin{eqnarray}
I_{\rm L} 
&=& \eminus \ N^{\rm min}_{\rm LM}\ \Big( G_{\rm M}\left(\eps_{\rm 0L}\right) - G_{\rm M}\left(\eps_{\rm 1L}\right)
\ +\ G_{\rm L}\left(\eps_{\rm 1L}\right) - G_{\rm L}\left(\eps_{\rm 0L}\right) \Big), \qquad
\label{Eq:I_L-boxcar-result}
\\
I_{\rm R} 
&=& \eminus \  N^{\rm min}_{\rm RM}\ \Big( G_{\rm M}\left(\eps_{\rm 1R}\right) - G_{\rm M}\left(\eps_{\rm 0R}\right)
\ +\ G_{\rm R}\left(\eps_{\rm 0R}\right) - G_{\rm R}\left(\eps_{\rm 1R}\right) \Big) ,\qquad
\label{Eq:I_R-boxcar-result}
\end{eqnarray}
with $I_{\rm M} = -I_{\rm L}-I_{\rm R}$.
Remember that $\partial_{\rm R}$ is a derivative with respect to $V_{\rm R}$ for fixed $\eps_{0i}$ and  $\eps_{1i}$, 
and the only $V_{\rm R}$ dependence is in $G_{\rm R}(\eps)$, we use  Eq.~(\ref{Eq:useful-derivative1}) 
to get 
\begin{eqnarray}
\partial_{\rm R} I_{\rm R} &=& 
{(\eminus)^2 \over h}\ N^{\rm min}_{\rm RM} \  \Big( f_{\rm R}\left(\eps_{\rm 0R}\right) - f_{\rm R}\left(\eps_{\rm 1R}\right) \Big) ,
\qquad
\end{eqnarray}
with $\partial_{\rm R} I_{\rm L}= 0$ and 
$\partial_{\rm R} I_{\rm M} = -\partial_{\rm R} I_{\rm R}$.
Similarly, the only $V_{\rm L}$ dependence is in $G_{\rm L}(\eps)$, hence
\begin{eqnarray}
\partial_{\rm L}  I_{\rm L} 
&=& {(\eminus)^2 \over h} \ N^{\rm min}_{\rm LM} \ \left[ f_{\rm L}\left(\eps_{\rm 1L}\right)- f_{\rm L}\left(\eps_{\rm 0L}\right) \right] , \qquad
\end{eqnarray}
with $\partial_{\rm L}  I_{\rm R} = 0$ and 
$\partial_{\rm L}  I_{\rm M} = -\partial_{\rm L} I_{\rm L}$.
Then $\partial_{\rm L} P_{\rm gen} = - I_{\rm L} - V_{\rm L}\partial_{\rm L} I_{\rm L}$ and $\partial_{\rm R} P_{\rm gen} = - I_{\rm R} -V_{\rm R}\partial_{\rm R} I_{\rm R}$.

The two contributions to the  heat-current out of reservoir M, defined above Eq.~(\ref{Eq:J-parts}), are
\begin{eqnarray}
J_{\rm M}^{\rm (L)}
\!\!\!   &=& \!\!  N^{\rm min}_{\rm LM} \big(F_{\rm M}\left(\eps_{\rm 1L}\right) - F_{\rm M}\left(\eps_{\rm 0L}\right) - F_{\rm L}\left(\eps_{\rm 1L}\right) + F_{\rm L}\left(\eps_{\rm 0L}\right) \! \big) ,
\nonumber \\
J_{\rm M}^{\rm (R)}
\!\!\!   &=& \!\!  
N^{\rm min}_{\rm RM} \big(F_{\rm M}\left(\eps_{\rm 0R}\right) - F_{\rm M}\left(\eps_{\rm 1R}\right) - F_{\rm R}\left(\eps_{\rm 0R}\right) + F_{\rm R}\left(\eps_{\rm 1R}\right) \! \big) .
\nonumber
\end{eqnarray}
Using  
Eq.~(\ref{Eq:useful-derivative2}), we get 
\begin{eqnarray}
\partial_{\rm R} J_{\rm M} \!\!&=&  \eminus \,N^{\rm min}_{\rm RM}\,   \Big[ \,G_{\rm R}\left(\eps_{\rm 1R}\right)-
G_{\rm R}\left(\eps_{\rm 0R}\right) 
\ + \ {\eps_{\rm 1R} \over h}f_{\rm R}\left(\eps_{\rm 1R}\right) -
 {\eps_{\rm 0R} \over h}f_{\rm R}\left(\eps_{\rm 0R}\right) \Big] ,
 \qquad
 \\
\partial_{\rm L} J_{\rm M} 
\!\!&=&\eminus\, N^{\rm min}_{\rm LM}\,  \Big[\, G_{\rm L}\left(\eps_{\rm 0L}\right) - G_{\rm L}\left(\eps_{\rm 1L}\right)
\ +\ {\eps_{\rm 0L} \over h} f_{\rm L}\left(\eps_{\rm 0L}\right) - {\eps_{\rm 1L} \over h} f_{\rm L}\left(\eps_{\rm 1L}\right) \Big] .
\qquad
\end{eqnarray}

\section{Useful derivatives and limits}

For any function $g(x)$
\begin{eqnarray}
{\rmd \over \rmd V_i} \int_{\eps_0}^{\eps_1} {\rmd \eps \over h} \,g\!\left({\eps-\eminus V_i \over \kB T_i}\right)
\ =\  
-{\eminus \over h} \left[g(x_1) -g(x_0)\right] 
\nonumber
\end{eqnarray}
where we defined $x_\alpha(V_i)=(\eps_\alpha-\eminus V_i)/(\kB T_i)$ for $\alpha =0,1$,
and use the fact that $V_i$ only appears in these limits on the integral.
Thus, for $G_j(\eps)$ in Eq.~(\ref{Eq:G_j+F_j})
we have
\begin{eqnarray}
{\rmd \over \rmd V_i} G_j(\eps) &=& {\eminus \over h} f_j(\eps) ,
\label{Eq:useful-derivative1}
\end{eqnarray}
Similarly for $F_j(\eps)$ in Eq.~(\ref{Eq:G_j+F_j}),
we have 
\begin{eqnarray}
{\rmd \over \rmd V_j} F_j(\eps) 
&=&
\kB T_j {\rmd \over \rmd V_j} \int_{\eps}^{\infty} {\rmd \tilde\eps \over h} \ \left({\tilde\eps-\eminus V_j\over \kB T_j}\right) \  f_j(\tilde\eps)
 + {\rmd \over \rmd V_j} \left[\eminus V_j  \int_{\eps}^{\infty} {\rmd \tilde\eps \over h} \  f_j(\tilde\eps) \right]
\nonumber \\
&=& 
\eminus  \left( G_j(\eps)+ {\eps \over h} \ f_j(\eps)  \right) .
\label{Eq:useful-derivative2}
\end{eqnarray}

Finally, we mention the limits of the dilogarithm functions that appear in $F_j(\eps)$.
The series expansion of the dilogarithm at small $z$ is 
${\rm Li}_2(z) = \sum_{n=1}^\infty n^{-2} z^n$.
One can then  extract the behaviour at $z=-\e^x$ for large $x$ using the equality
${\rm Li}_2(-\e^x) +  {\rm Li}_2(-\e^{-x}) = -{\pi^2\big/6} - {x^2 \big/ 2}$.
Inserting the above small $z$ expansion into this, gives 
\begin{eqnarray}
 {\rm Li}_2(-\e^x) 
&=& - {x^2 \over 2} -{\pi^2 \over 6} -  \sum_{n=1}^\infty {(-1)^n  \over n^2} \,\e^{-nx} \, .
\label{Eq:dilog-large-x}
\end{eqnarray}


\vskip 20mm


\begin{thebibliography}{1}




\bibitem{whitney-prl2014}
R.S.\ Whitney,
``Most efficient quantum thermoelectric at finite power output'',
Phys.\ Rev.\ Lett.\ {\bf 112}, 130601 (2014).

\bibitem{whitney2015}
R.S.\ Whitney,
``Finding the quantum thermoelectric with maximal efficiency and minimal entropy production at given power output'',
Phys.\ Rev.\ B {\bf 91}, 115425 (2015).


\bibitem{Entin2010}
O. Entin-Wohlman, Y. Imry, A. Aharony, 
``Three-terminal thermoelectric transport through a molecular junction'',
Phys.\ Rev.\ B {\bf 82}, 115314 (2010).
\bibitem{Sanchez2011}
R.~S\'anchez, M.~B\"uttiker
``Optimal energy quanta to current conversion'',
Phys.\ Rev.\ B {\bf 83}, 085428 (2011).
\bibitem{Sothmann2012a}
B.\ Sothmann, R.\ S\'anchez, A.N.\ Jordan, M.\ B\"uttiker,
``Rectification of thermal fluctuations in a chaotic cavity heat engine''
Phys.\ Rev.\ B {\bf 85}, 205301 (2012).
\bibitem{Entin2012}
O. Entin-Wohlman, A. Aharony,
``Three-terminal thermoelectric transport through a molecule placed on an Aharonov-Bohm ring'',
Phys.\ Rev.\ B {\bf 85}, 085401 (2012).
\bibitem{Jiang2012}
Jian-Hua Jiang, Ora Entin-Wohlman, Yoseph Imry, 
``Thermoelectric three-terminal hopping transport through one-dimensional nanosystems'',
Phys.\ Rev.\ B {\bf 85}, 075412 (2012).
\bibitem{Horvat2012}
M.\ Horvat, T.\ Prosen, G.\ Benenti, G.\  Casati, 
``Railway switch transport model'',
Phys.\ Rev.\ E {\bf 86}, 052102 (2012).
\bibitem{Brandner2013}
K.\ Brandner, K.\ Saito, U.\ Seifert, 
``Strong bounds on Onsager coefficients and efficiency for three terminal thermoelectric transport in a magnetic field'',
Phys.\ Rev.\ Lett.\ {\bf 110}, 070603 (2013).
\bibitem{Balachandran2013}
V.\  Balachandran, G.\  Benenti, G.\  Casati,
``Efficiency of three-terminal thermoelectric transport under broken-time reversal symmetry'',
Phys.\ Rev.\ B {\bf 87}, 165419 (2013).
\bibitem{Jiang2013}
J.-H.\ Jiang, O.\ Entin-Wohlman, Y.\  Imry,
``Three-terminal semiconductor junction thermoelectric devices: improving performance'',
New J.\ Phys.\ {\bf 15}, 075021 (2013).
\bibitem{Entin-Wohlman2013}
O. Entin-Wohlman, A. Aharony, Y. Imry,
``Mesoscopic Aharonov-Bohm Interferometers:Decoherence and Thermoelectric Transport'',
{\it In Memory of Akira Tonomura: Physicist and Electron Microscopist}, 
Eds. Kazuo Fujikawa and Yoshimasa A. Ono (World Scientific, Singapore, 2013).
\bibitem{Sanchez2013}
R.\ S\'anchez, B.\  Sothmann, A.N.\ Jordan, M.\  B\"uttiker,
``Correlations of heat and charge currents in quantum-dot thermoelectric engines'',
New J.\ Phys.\ {\bf 15}, 125001 (2013).
\bibitem{Jiang2014}
J.-H.\ Jiang, 
``Enhancing efficiency and power of quantum-dots resonant tunneling thermoelectrics in three-terminal geometry by cooperative effects'',
J.\ Appl.\ Phys. {\bf 116}, 194303 (2014).
\bibitem{Mazza2014}
F.\ Mazza, R.\ Bosisio, G.\ Benenti, V.\ Giovannetti, R.\ Fazio, and F.\ Taddei,
``Thermoelectric efficiency of three-terminal quantum thermal machines'',
New J.~ Phys.\ {\bf 16}, 085001 (2014).
\bibitem{Mazza2015}
F.\ Mazza, S.\ Valentini, R.\ Bosisio, G.\ Benenti, V.\ Giovannetti, R.\ Fazio, and F.\ Taddei,
``Separation of heat and charge currents for boosted thermoelectric conversion'',
Phys.\ Rev.\ B {\bf 91}, 245435 (2015).
\bibitem{Sothmann2015}
P.P.\ Hofer, and B.\ Sothmann, 
``Quantum heat engines based on electronic Mach-Zehnder interferometers'',
Phys.\ Rev.\ B {\bf 91}, 195406 (2015).
\bibitem{Sanchez2015a-qu-hall}
R.\ S\'anchez, B.\ Sothmann, and A.N.\ Jordan,
``Chiral thermoelectrics with quantum Hall edge states'',
Phys.\ Rev.\ Lett.\ {\bf 114}, 146801 (2015)
 \bibitem{Sanchez2015b-qu-hall}
R.\ S\'anchez, B.\ Sothmann, and A.N. Jordan,
``Effect of incoherent scattering on three-terminal quantum Hall thermoelectrics'',
Physica E {\bf 75}, 86 (2016).
\bibitem{Jiang2015}
J.-H.\ Jiang, B.\ Kumar Agarwalla, D.\ Segal,
``Efficiency Statistics and Bounds of Time-Reversal Symmetry Broken Systems'',
 Phys.\ Rev.\ Lett.\ {\bf 115}, 040601 (2015).


\bibitem{Roche15}
B.\ Roche, P.\ Roulleau, T.\ Jullien, Y.\ Jompol, I.\ Farrer, D.A.\ Ritchie and D.C.\ Glattli, Nature Comm. {\bf 6}, 6738 (2015).
\bibitem{Hartmann15}
F. Hartmann, P. Pfeffer, S. H\"ofling, M. Kamp and L. Worschech, Phys. Rev. Lett. {\bf 114},  146805 (2015)
\bibitem{Thierschmann15}
H. Thierschmann, R. S\'anchez, B. Sothmann, F. Arnold, C. Heyn, W. Hansen, H. Buhmann, L. W. Molenkamp, Nature Nanotech., in press (2015); doi: 10.1038/nnano.2015.176


\bibitem{books}
H.J.~Goldsmid, {\it Introduction to Thermoelectricity} (Springer, Heidelberg, 2009). 
\bibitem{DiSalvo-review}
F.J.~DiSalvo, ``Thermoelectric Cooling and Power Generation'',
Science {\bf 285}, 703 (1999).
\bibitem{Shakouri-reviews}
A.~Shakouri and M.~Zebarjadi, 
``Nanoengineered Materials for Thermoelectric Energy Conversion'',
Chapt 9 of {\it Thermal nanosystems and nanomaterials}, S.~Volz (Ed.)  (Springer, Heidelberg, 2009).
A.~Shakouri,
``Recent Developments in Semiconductor Thermoelectric Physics and Materials'', 
Annu. Rev. Mater. Res. {\bf 41}, 399 (2011).


\bibitem{Christen-Buttiker1996a}
T.\ Christen, and M.\ B\"uttiker, 
``Gauge invariant nonlinear electric transport in mesoscopic conductors.''
Europhys.\ Lett.\ 35, 523 (1996). 


\bibitem{Jordan-Sothmann-Sanchez-Buttiker2013}
A.N.\ Jordan, B.\ Sothmann, R.\ Sanchez, and M.\ B\"uttiker,
``Powerful and efficient energy harvester with resonant-tunneling quantum dots'',
Phys.~Rev.~B, {\bf 87}, 075312 (2013). 
\bibitem{Sothmann-Sanchez-Jordan-Buttiker2013}
 B.\ Sothmann, R.\ Sanchez, A.N.\ Jordan,and M.\ B\"uttiker,
 ``Powerful energy harvester based on resonant-tunneling quantum wells'',
New J.\ Phys.\ {\bf 15} (2013) 095021. 



\bibitem{Bekenstein}
J.D.\ Bekenstein, 
``Energy Cost of Information Transfer'',
Phys.\ Rev.\ Lett.\ {\bf 46}, 623 (1981).
J.D.\ Bekenstein, 
``Entropy content and information flow in systems with limited energy'',
Phys.\ Rev.\ D {\bf 30}, 1669 (1984).
\bibitem{Pendry1983}
J.B.~Pendry, 
``Quantum limits on the flow of information and entropy'',
J.~Phys.~A.: Math.\ Gen.\ {\bf 16}, 2161 (1983).
\bibitem{Molenkamp-Peltier-thermalcond}
L.W.~Molenkamp, Th.\ Gravier, H.\ van Houten, O.J.A.\ Buijk, M.A.A.\ Mabesoone, C.T.\ Foxon, 
``Peltier coefficient and thermal conductance of a quantum point contact'',
Phys.\ Rev.\ Lett.\ {\bf 68}, 3765 (1992).
\bibitem{Jezouin2013}
S.\ Jezouin, F.\ Parmentier, A.\ Anthore, U.\ Gennser, A.\ Cavanna, Y.\ Jin, and F.\ Pierre, 
``Quantum limit of heat flow across a single electronic channel'', 
Science {\bf 342}, 601 (2013).




\bibitem{Curzon-Ahlborn1975}
F.L.~Curzon, and B.~Ahlborn, 
``Efficiency of a Carnot engine at maximum power output'',
Am.~J.~Phys.\  {\bf 43}, 22 (1975).

\bibitem{Yvon1956}
J.\ Yvon, Proceedings of the International Conference on Peaceful Uses of Atomic Energy (Vol.~2), 
``Saclay Reactor: acquired knowledge by two years experience in heat transfer using compressed gas'', 
p.~337 (United Nations, New York, 1956). 
\bibitem{Chambadal57}
P. Chambadal, Les Centrales Nucl\'eaires (Armand Colin, 1957), p.~41.
\bibitem{Novikov57}
 I. I. Novikov, ``The Efficiency of Atomic Power Stations'', J.\ Nucl.\ Energy II 7, 125 (1958) [Atomnaya Energiya 3, 409 (1957)].

 
\bibitem{Casati-review}
G.~Benenti, G.~Casati, T.~Prosen, and K.~Saito,
``Colloquium: Fundamental aspects of steady state heat to work conversion'',
Eprint arXiv:1311.4430.

\bibitem{Benenti-et-al2011}
G.\ Benenti, K.\ Saito, and G.\ Casati,
``Thermodynamic Bounds on Efficiency for Systems with Broken Time-Reversal Symmetry'',
Phys.\ Rev.\ Lett.\ {\bf 106}, 230602 (2011).

\bibitem{Entin-Jiang-Imry2014}
O.\ Entin-Wohlman, J.-H.\ Jiang, and Y.\ Imry,
``Efficiency and dissipation in a two-terminal thermoelectric junction, emphasizing small dissipation''.
Phys.\ Rev.\ E {\bf 89}, 012123 (2014).


\bibitem{Mahan-Sofo1996}
G.D.~Mahan, and J.O.~Sofo, 
``The best thermoelectric'', 
Proc.\ Nat.\ Acad.\ Sci.\ USA, {\bf 93}, 7436 (1996).

\bibitem{Humphry-Newbury-Taylor-Linke2002}
T.E.~Humphrey, R.\ Newbury, R.P.\ Taylor, H.\ Linke,
``Reversible quantum Brownian heat engines for electrons'',
Phys.\ Rev.\ Lett.\ {\bf 89}, 116801 (2002). 

\bibitem{Humphrey-Linke2005}
T.E.~Humphrey, and H.~Linke, 
``Reversible Thermoelectric Nanomaterials'',
Phys.\ Rev.\ Lett.\ {\bf 94}, 096601 (2005).


\bibitem{Brandner2015}
K.~Brandner, U.~Seifert,
``Bound on Thermoelectric Power in a Magnetic Field within Linear Response'',
Phys.\ Rev.\ E {\bf 91}, 012121 (2015). 

\bibitem{Enquist-Anderson1981}
H.- L.\ Engquist and P.W.\ Anderson. 
``Definition and measurement of the electrical and thermal resistances'', 
Phys.\ Rev.\ B {\bf 24}, 1151(R) (1981).
\bibitem{Buttiker1986-4probe}
M.\ B\"uttiker, ``Four-Terminal Phase-Coherent Conductance'', 
Phys.\ Rev.\ Lett.\ {\bf 57}, 1761 (1986).
\bibitem{voltage-probe}
M.~B\"uttiker, 
``Coherent and sequential tunneling in series barriers'',
IBM J.\ Res.\ Dev.\ {\bf 32}, 63 (1988).
\bibitem{Imry-book}
Y.\ Imry, ``Introduction to Mesoscopic Physics'' (Oxford University Press, Oxford, 2002).




\bibitem{pjw2007}
C.\ Petitjean, Ph.\ Jacquod, R.S.\ Whitney
``Dephasing in the semiclassical limit is system-dependent",
JETP Letters, 86, 647 (2007).
\bibitem{wjp2008}
R.S.\ Whitney, Ph.\ Jacquod, C.\ Petitjean
``Dephasing in quantum chaotic transport: a semiclassical approach'', 
Phys.\ Rev.\ B {\bf 77}, 045315 (2008). 


\bibitem{Nenciu2007}
G.\ Nenciu,
``Independent electron model for open quantum systems: Landauer-B\"uttiker formula and strict positivity of the entropy production'',
J.\ Math.\ Phys.\ {\bf 48}, 033302 (2007).

\bibitem{whitney2013}
R.S.~Whitney, 
`` Thermodynamic and quantum bounds on nonlinear DC thermoelectric transport.'' 
Phys.\ Rev.\ B {\bf 87}, 115404 (2013).


\bibitem{Cleuren2012}
See e.g.\ Eq.~(11) of 
B.\ Cleuren, B.\ Rutten, and C.\ Van den Broeck,
``Cooling by Heating: Refrigeration Powered by Photons'',
Phys.\ Rev.\ Lett.\ {\bf 108}, 120603 (2012). 

\bibitem{wshs2016}
R.S. Whitney, R.\ S\'anchez, F.\ Haupt, and J.\ Splettstoesser,
``Thermoelectricity without absorbing energy from the heat sources'',
Physica E {\bf 75}, 257 (2016).







\end{thebibliography}
\end{document}